\documentclass[pdf-latex,sn-mathphys-num]{sn-jnl}% Math and Physical Sciences Numbered Reference Style
\usepackage{graphicx}%
\usepackage{multirow}%
\usepackage{amsmath,amssymb,amsfonts}%
\usepackage{amsthm}%
\usepackage{mathrsfs}%
\usepackage[title]{appendix}%
\usepackage{xcolor}%
\usepackage{textcomp}%
\usepackage{manyfoot}%
\usepackage{booktabs}%
\usepackage{algorithm}%
\usepackage{algorithmicx}%
\usepackage{algpseudocode}%
\usepackage{listings}%
\usepackage{ccaption}
\usepackage{makecell}
\usepackage{placeins}
\usepackage{float}
\usepackage{capt-of}
\theoremstyle{thmstyleone}%
\theoremstyle{thmstyletwo}%

\theoremstyle{thmstylethree}%

\begin{document}

\title[Article Title]{Domain Boundaries in a Metallic Distortive Polar Metal }


\author*[1]{\fnm{A.} \sur{Savovici}}\email{a.savovici@mpi-susmat.de}
\equalcont{These authors contributed equally to this work.}

\author*[1]{\fnm{B.} \sur{Ratzker}}\email{b.ratzker@mpi-susmat.de}
\equalcont{These authors contributed equally to this work.}

\author[1]{\fnm{X.} \sur{Zhou}}
\author[1]{\fnm{S.} \sur{Zaefferer}}
\author[1]{\fnm{M.} \sur{Ruffino}}
\author[2]{\fnm{I.} \sur{Radulov}}
\author[1]{\fnm{P.} \sur{Jovičević-Klug}}
\author[1]{\fnm{S.} \sur{Katnagallu}}
\author[1]{\fnm{A.} \sur{Hamzehei}}
\author[1]{\fnm{P.} \sur{Watermayer}}
\author[1]{\fnm{A.} \sur{Vogel}}
\author[1]{\fnm{J.} \sur{Neugebauer}}
\author[1]{\fnm{M.} \sur{Jovičević-Klug}}
\author[1]{\fnm{C.} \sur{Freysoldt}}
\author*[1]{\fnm{D.} \sur{Raabe}}\email{d.raabe@mpi-susmat.de}

\affil[1]{\orgname{Max Planck Institute for Sustainable Materials}, \orgaddress{\street{Max-Planck-Str. 1}, \city{Düsseldorf}, \postcode{40237}, \country{Germany}}}

\affil[2]{\orgname{Fraunhofer Research Institution for Materials Recycling and Resource Strategies IWKS}, \orgaddress{\street{Aschaffenburger Str. 121}, \city{Hanau}, \postcode{63457}, \country{Germany}}}


\abstract{Polar metals are an underexplored material class combining two properties that are typically incompatible, namely a polar crystal structure and reasonable electrical conductivity. These intriguing materials offer a unique combination of properties, potentially relevant to optoelectronics, catalysis, memory devices, among other applications. The distortive polar metal (DPM) subclass forms through a symmetry-lifting phase transformation into a non-centrosymmetric polar crystal structure. In the process, domains with uniform geometric polar directions form, oftentimes separated by domain boundaries with polarity discontinuities arranged in “charged” head-to-head (H-H) or tail-to-tail (T-T) morphologies. To date, only metallic oxide DPM microstructures have been studied. Here we reveal in the intermetallic DPM Mn$_5$Al$_8$ different  surface interactions and electron transfer reactivity at domain boundaries depending on their H-H or T-T character. Variable surface reactivity suggests localized changes in electronic work functions due to an increase (H-H) or decrease (T-T) in the electronic density of states.  These findings suggest that metallic DPMs may offer functionalizable domain boundaries and deserve increased attention given that they allow tunable chemistries and various thermomechanical processing or transformation protocols. Ultimately, this study proposes unconventional metal physics, propelling the discovery and design of advanced electronic materials and devices. }


\maketitle

\section{Introduction}\label{sec:main}
Non-centrosymmetric crystals exhibit multiple functional properties such as piezoelectricity, ferroelectricity, and pyroelectricity. Of the 21 crystal classes which lack inversion symmetry, 10 allow the existence of a permanent dipole moment and are commonly termed polar \cite{ITC, Ok2006}. Ferroelectric materials are typically insulators that undergo a centrosymmetry-reducing transition below their Curie temperature, exhibiting spontaneous polarization along an axis whose direction can be switched via an applied electric field. Polar metals on the other hand are defined by the coexistence of a polar crystal structure and sufficient mobile charge carriers enabling intrinsically robust electrical conductivity \cite{bhowal2023}. Since electric fields are screened by itinerant electrons in polar metals, the direction of the geometric polarity directions cannot be reversed by an applied electric field. A ‘ferroelectric transition’ in a metal was first discussed by Anderson and Blount \cite{Anderson1965} in 1965, which served as the nucleus for the polar metal field, where they envisioned the coexistence of polarity and metallicity as a decoupling of the centrosymmtery-breaking and free-carrier generating elements. Recent interest in the polar metal field was incited by the direct observation of an Anderson and Blount-style second-order transition in the metallic oxide LiOs$\rm{O}_3$ \cite{Shi2013} in 2013. Since then, the field has expanded and the term ‘polar metal’ now refers not only to materials that undergo ``ferroelectric-like” transitions, but to a broad material class with combined polarity and metallicity, spanning diverse chemistries, structures, and morphologies \cite{bhowal2023,hockoxyoung2023}.

The polar metal subclass criteria, as given by Hickox et al. \cite{hockoxyoung2023}, considers key factors that differentiate polar metals – namely, whether the polar metal state occurs following a structural transformation, or whether low dimensionality \cite{ke2021}, doping \cite{zhao2018,Gu2017}, or heterointerface phenomena \cite{Cao2018} govern the coexistence of polarity and metallicity. The distortive polar metal (DPM) subclass is defined by a first- or second-order centrosymmetry-lifting transformation which can exhibit an order/disorder or displacive quality, or a mixture of both. Broadly speaking, the engineering design guide for DPMs is the ‘weak coupling principle’ \cite{bhowal2023}, where the electronic structure at the Fermi surface is primarily derived from orbital states that are de-coupled from the ions undergoing polar distortions. This behavior is exhibited in ABO$_3$ and Ruddelsden-Popper perovskite LiOsO$_3$ \cite{Shi2013}, Ca$_3$Ru$_2$O$_7$ \cite{Lei2018} polar metals respectively. An analogous principle exists in the layered van der Waals structure WTe$_2$ \cite{Fei2018}, where a ‘decoupled space mechanism’ \cite{bhowal2023} between transversally oriented polarization and conductivity axes define the polar metal state. Overall, the absence of centrosymmetry in polar metals enables several unique physical effects, which in combination with additional material-specific properties has led to suggested applications such as electric-field control of magnetism \cite{Jager2024}, magnetopiezoelectricity \cite{Varjas2016}, non-linear optics \cite{Padmanabhan2018}, thermopower \cite{Puggioni2014}, ohmic contact switching \cite{Hwang2025}, and topological-driven properties \cite{bhowal2023}.

DPMs undergo a symmetry reducing transformation, where polar domains form to minimize energies oftentimes related to ferroic order parameters, yielding a functionalizable domain microstructure \cite{Padmanabhan2018}. As a polar material, each domain is defined by a uniform crystallographic polarization direction ($\eta$), that encodes the orientation of the polar axis in 3D and the signed amplitude of the geometric polar distortion. The region separating two domains exhibiting dissimilar polar directions is typically called a ‘domain boundary’ \cite{Catalan2012}. Discontinuities of polar direction between two domains $\eta_1$, $\eta_2$ leads to “bound charge”, and can be expressed as: $\sigma_P=(\eta_2-\eta_1)\cdot\mathbf{n}_1$, where $\sigma_P$ is the surface “bound charge” density and $\mathbf{n}_1$ is the wall normal unit vector pointing into domain 1. The term “charge” is somewhat unfortunate in DPMs, becuase it could be mistaken for implying a net electric charge at the boundary, which cannot exist inside a metal. Yet, we follow previous literature \cite{Stone2019,hockoxyoung2023} for lack of a better term. In ferroelectrics, bound charge generates an energetically unfavorable de-polarizing electric field, causing free carriers (electrons, holes, or mobile ions) to aggregate at charged domain boundaries, screening the field \cite{Bednyakov2018}. Both convergent head-to-head (H-H) or divergent tail-to-tail (T-T) charged domain boundaries may form between adjacent domains. In DPMs “charged” H-H and T-T boundaries still form \cite{Stone2019}, despite electrical polarization in a metal being ill-defined. Mechanistically, “charged” domain boundaries in a DPM should differ from ferroelectrics since electric fields are screened in a metal. Nevertheless, “bound charge” and “charged” domain boundaries still exist as concepts for DPMs, describing the structural polar reorientations across domain boundaries and the resulting screening response of the itinerant electrons \cite{Stone2019}. Domain boundaries are known to exhibit intriguing properties, vastly different from the bulk. For example, ferroelectric domain walls exhibit enhanced conductivity \cite{Seidel2010}, superconductivity \cite{Aird1998}, photovoltaic effect \cite{Yang2010}, making them functionalizable, technologically relevant, and extensively researched \cite{Catalan2012,Seidel2020}. In contrast, although polar metal domain boundaries can also exhibit exotic behavior \cite{Stone2019}, they have received surprisingly limited attention and are poorly understood. 
\textit{Metallic} DPMs and their domain boundaries have been essentially overlooked – despite their mention in recent polar metal reviews \cite{bhowal2023,hockoxyoung2023}. This point is emphasized to distinguish metallic DPMs made of strictly metallic species from the routinely studied metallic \textit{oxide} DPMs (LiOsO$_3$ \cite{Shi2013}, Ca$_3$Ru$_2$O$_7$ \cite{Lei2018}). It is plausible that metallic polar metals have received such limited attention since their apparent simplicity should at first glance forbid remarkable properties. However, it is precisely their incongruent combination of simple chemistry and complex structure that may have allowed them to hide rich, unexplored physics right in plain sight.
In this study, we set out to analyze domain boundaries in a metallic DPM – Mn$_5$Al$_8$. The symmetry-reducing transformation creates a herringbone morphology with ``charged” H-H/T-T domain boundaries, visualized with advanced electron microscopy characterization techniques. Direct and indirect surface probing of domain boundary types revealed variable surface reactivity, where a surplus and depletion of free carriers for H-H/T-T respectively is suggested. While the finding of ``charged” domain boundaries is in general agreement with polar metal theory \cite{Gu2023}, it has not been observed or considered. We hypothesize that the complex phenomena at domain boundaries originates from polar discontinuities, something common in insulating ferroelectrics – but unexpected for a simple metal.

\section{Results}
\subsection{Structural transition in Mn$_5$Al$_8$}

\begin{figure}
    \centering
    \includegraphics[width=\textwidth]{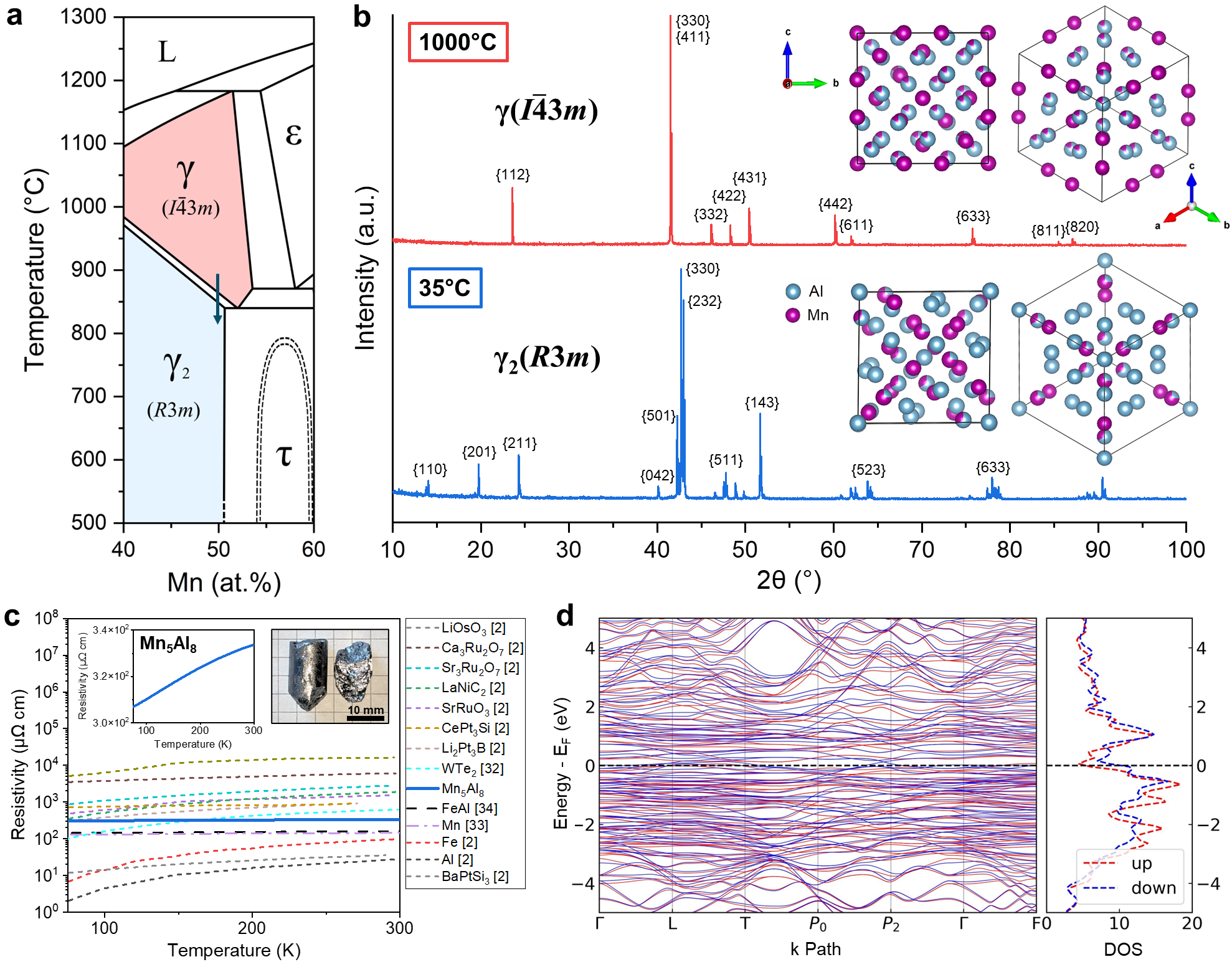}
    \caption{\textbf{Structural and electronic assessment of polar metal state.} \textbf{a,} Mn-Al phase diagram showing suggested high-temperature (cubic, $\gamma$-brass, red) and low temperature (rhombohedral, $\gamma_2$, blue) phases. The arrow in this diagram highlights the composition chosen and transformation pathway. \textbf{b,} Room temperature and in-situ $1000^{\circ}$C high temperature XRD data of the cubic $\gamma_2$–brass and polar $\gamma_2$ phase, with the cubic/pseudo-cubic respective unit cells viewed along $[100]$ and $[111]$. \textbf{c,} Electrical resistivity vs temperature data of Mn$_5$Al$_8$ is plotted alongside many studied polar metal compounds of different classifications \cite{hockoxyoung2023,Ali2014} and some metals \cite{hockoxyoung2023,Desai1984,ZHANG2001}, and the left inset confirms positive correlation between resistivity and temperature. The inset on the right shows a photograph of as-cast sample pieces. \textbf{d,} The calculated electronic band structure and density of states are given, confirming metallic behavior.}
    \label{fig:figure1}
\end{figure}

The symmetry reducing transition from cubic ($\gamma$) $\rightarrow$ rhombohedral ($\gamma_2$) in off-stoichiometric composition Mn$_{51}$Al$_{49}$ was observed to occur at $\sim$$880^{\circ}$C, where both high and low temperature phases were captured with elevated temperature X-ray diffraction (XRD). Diffractograms were captured at $20^{\circ}$C intervals between $600^{\circ}$C to $1000^{\circ}$C, where the full XRD data set is given in Supplementary Information (SI) Fig. S1a. The critical transition temperature as given by the phase diagram is in general agreement with differential scanning calorimetry (DSC) measurements showing it is a first-order phase transformation (Fig. S1b). The relevant portion of the phase diagram is given in Fig. \ref{fig:figure1}a, showing the high temperature $\gamma$-brass phase. Previous measurements \cite{Ellner1990} and theory \cite{Liu1999} suggest either an A2 or B2 (Strukturbericht) phase field above the Mn-rich $\gamma_2$ region - whereas we observe a BCC-related cubic $\gamma$-brass structure ($I\bar{4}3m$  space group) as shown in Fig. \ref{fig:figure1}b. BCC and $\gamma$-brass are strongly related where $\gamma$-brass can be envisioned as a $3\times3\times3$ BCC supercell with 2 atoms removed \cite{bradley1926}, forcing small atomic shuffles (due to change in atom count) allowing reflections in $\gamma$-brass not attributable to A2 or B2. Nevertheless, identification of the high temperature structure is non-trivial, as many reports strongly suggests $m\bar{3}m$ point-group symmetry\cite{Zeng2018} (e.g. B2), and differentiation between similar phases (e.g. Ir$_{3}$Ge$_{7}$ structure\cite{Swenson97}) is frustrated. The low temperature rhombohedral ($\gamma_2$) phase is identified by XRD (as given in Fig. \ref{fig:figure1}b), and confirms the polar structure ($R3m$ space group belongs to one of the ten polar crystal classes). Both $\gamma$-brass and $\gamma_2$ unit cells are shown alongside the high/low temperature diffractograms in Fig. \ref{fig:figure1}b, where $\gamma_2$ is displayed in its pseudo-cubic (PC) body-centered rhombohedral unit cell \cite{Zeng2018} (Extended Data Fig. 2 displays the relationship between primitive rhombohedral, hexagonal-R, and body-centered rhombohedral). The cubic $\gamma$-brass phase has a lattice constant $a=9.218$ \AA, whereas $\gamma_2$ in the pseudo-cubic unit cell exhibits $a=8.99374$ $\AA$ with $\alpha=88.943^{\circ}$. Since the trigonal unit cell hosts 84 atoms, the displacive transformation is more intuitively visualized with cubic ($\gamma$-brass) and pseudo-cubic unit cells ($\gamma_2$), where site shifts between the central 26-atom $\gamma$-cluster can be easily observed (see Fig. S2). 

While the polar structure is confirmed by the XRD structural measurements, the metallic designation is given by temperature-dependent resistivity of Mn$_5$Al$_8$ as shown in Fig. \ref{fig:figure1}c (blue line) and the electronic density of states (DOS) presented in Fig. \ref{fig:figure1}d. The electrical resistivity of Mn$_5$Al$_8$ is plotted alongside various polar metals \cite{hockoxyoung2023,Ali2014} in Fig. \ref{fig:figure1}c; elemental Mn \cite{Desai1984}, Fe, and Al are plotted as reference in addition to an Fe-Al compound \cite{ZHANG2001}. To note, the low variation of resistivity as a function of temperature is routinely observed in other alloys and intermetallics \cite{SHIRAI1995}. The inset of Fig. \ref{fig:figure1}c shows the resistivity of our off-stoichiometric Mn$_{51}$Al$_{49}$ sample plotted without logarithmic scale, showing a clear positive correlation between resistivity and temperature ($\frac{d\rho}{dt}>0$), which allows adequate metallic classification. Additionally, the calculated electronic dispersion curves and DOS from density functional theory (DFT) for the stoichiometric Mn$_5$Al$_8$ phase is shown in Fig. \ref{fig:figure1}d. The non-zero DOS near the Fermi level (dashed line) again confirms metallic behavior and is in agreement with previous reports \cite{Thimmaiah2017}. The identification of both a polar crystal class ($R3m$ space group) and metallic behavior (resistivity, electronic band structure) adequately describe a polar metal state by the accepted definitions within the literature \cite{bhowal2023,hockoxyoung2023,hickoxyoung2021}.

The polar/structural transformation ($\gamma\rightarrow\gamma_2$) is martensitic (i.e. displacive) and produces a microstructure rigorously defined as rank-2 laminate twins \cite{bhattacharya2003,Tsou2011}, which can form with diverse morphologies as given by compatible/nearly-compatible domain interface relationships \cite{liu2016} (see Fig. \ref{fig:figure2}a and Fig.S3). These domains form in accordance with elastic energy minimization as proposed by the constrained theory \cite{Ball1987}. The symmetry-reducing transformation $I\bar{4}3m$ $\rightarrow$ $R3m$ creates four crystallographic equivalent polar variants with polar axis along $[0001]_H || [111]_R || [111]_{PC}$ (parallel to the 3-fold rotation axis) \cite{bhowal2023,Erhart2004}, where subscript $H$, $R$, $PC$ correspond to Hexagonal, Rhombohedral, and Pseudo-Cubic respectively. The multi-rank laminate develops a morphology common for other strain-accommodating transformations (e.g. in shape-memory alloys \cite{Tsou2015}, thermoelectrics \cite{Lee2015,Vermeulen2016}, and ferroelectrics \cite{ricote1999}). Although multi-rank laminate twins can form in a dizzying array of patterns as shown by Liu and Tsou \cite{liu2016}, this study focuses on the ubiquitous herringbone morphology.

\subsection{Polar Domain Structure and Boundary Analysis}
The electron backscatter diffraction (EBSD) micrograph in Fig. \ref{fig:figure2}b shows four different orientation variants which are separated by twin interfaces (i.e., domain boundaries) of approximately $90^{\circ}$ misorientation about $\langle10\bar{1}2\rangle$ (black lines). As the pole figure (PF) in Fig. \ref{fig:figure2}b inset shows, the $\gamma_2$ domains are formed from a parent $\gamma$ grain with typical Cube-Rhombohedral relationships $[111]_C||[0001]_H$ and $(011)_C||(10\bar{1}0)_H$. The domain boundary traces of the mixed $\{100\}_{PC}$ and $\{011\}_{PC}$ twins are denoted by orange and white lines, respectively in Fig. \ref{fig:figure2}b. The domains exhibit minor orientation changes with significant orientation gradients of up to $\sim$$3^{\circ}$/$\mu$m localized at all twin boundaries regardless of twin plane. Fig. S3c shows this in the form of a grain reference orientation deviation (GROD) map. The fact that no dislocations have been observed even in the areas of highest orientation gradients this suggests that strong elastic strains localized at the domain boundaries persist in the sample likely caused by accommodation of the transformation strain among neighboring orientation variants. While each individual domain variant has a distinct polar axis, the direction is not articulated on these images (it is along $[0001]_H$ or $[111]_{PC}$). Though the $I\bar{4}3m$ $\rightarrow$ $R3m$ transformation produces variants with polar axes defining the regular tetrahedron of the parent cubic structure (i.e. $[111]$, $[\bar{1}\bar{1}1]$, $[\bar{1}1\bar{1}]$, $[1\bar{1}\bar{1}]$ or their inverse), we observe in this herringbone morphology polar axes not expected though reported before \cite{Sande1979}. The herringbone morphology exhibiting both $\{100\}_{PC}$ and $\{110\}_{PC}$ twin interfaces has been observed to relieve the most volume-change strain (rather than morphologies exhibiting pure $\{100\}_{PC}$ or pure $\{110\}_{PC}$ twins), though is not electrostatically neutral (in dielectrics) \cite{Vermeulen2016}. 

High-angle annular dark field scanning transmission electron microscopy (HAADF-STEM) with probe-corrected imaging allows characterization of the domain boundaries with atomic resolution. The high-resolution micrograph given in Fig. \ref{fig:figure2}c down a $[21\bar{3}\bar{2}]_H || [001]_{PC}$ zone axis shows a $\sim$$90^{\circ}$ boundary, in agreement with EBSD data (Fig. \ref{fig:figure2}b). The twinning interface is $(20\bar{2}1)_H$ or $(100)_{PC}$. This is only one of the two twin types present in the mixed twin herringbone morphologies i.e. $\{100\}_{PC}$ and $\{110\}_{PC}$. No secondary phases or ‘complexion phases’ \cite{Cantwell2014} are visible in the TEM analysis at the interface boundaries (further TEM micrographs are given in Fig. S4). The $\{100\}_{PC}/\{110\}_{PC}$ mixed-twin herringbone morphology allows the formation of head-to-head (H-H) and tail-to-tail (T-T) domain wall boundaries (polar axis discontinuities across the boundary), whose electrical properties are still poorly understood in polar metals. As true for pseudo-cubic trigonal phases, domain boundaries between adjacent domains can be either $109^{\circ}$ or $71^{\circ}$, where angles are formed between respective $\langle111\rangle$  polar vectors of individual domains \cite{Huang2011}. 

In efforts to better visualize polarity and polar domain boundaries, four-dimensional STEM differential phase contrast (4D-STEM DPC) measurements were performed (see Methods). 4D-STEM DPC allows for quantification of the deflection (spatially resolved phase shift) of the transmitted probe caused by local in-plane electric or magnetic fields \cite{Nordahl2024}, making it a powerful method for confirming polar domains in ferroelectrics \cite{Conroy2020,Takamoto2024}. Despite electric polarization being ill defined within a metal, and polar metal theory suggesting all internal electric fields be screened by free carriers, 4D-STEM DPC can still provide insight into geometric polarity and enable visualization of domain boundary character, as demonstrated in studies of ``charged” domain boundaries in Ca$_3$Ru$_2$O$_7$ \cite{lei2017}. Fig. \ref{fig:figure2}e displays the center-of-mass (CoM) vector-displacement map extracted from the 4D-STEM dataset with the pyDPC4D package \cite{Zhou2023}. This map reveals the four crystallographic variants that assemble into the characteristic herringbone morphology, consistent with the STEM micrographs shown in Extended Data Fig. 4. The color wheel indicates the direction of each displacement vector, while hue connotes its magnitude. While in principle, the vector magnitude should be proportional to local electric-field strength, the contrast pattern in Fig. \ref{fig:figure2}e is strongly influenced by diffraction effects (see Extended Data Fig. 5), since unscreened electric fields in a metal are not expected. See further offaxis measurements and simulations made with ReciPro\cite{ReciPro2022} software in Figs. S5 and Figs. S6. Intensity redistributions of Bragg disks and Kikuchi bands produced by structural differences between variants can modify the observed CoM and overlap with the rigid-body shifts of the bright-field Bragg disk typically seen in p-n junctions \cite{Shibata2015}. Diffraction contrast is common for ferroelectrics \cite{Takamoto2024,Strauch2023}, and highlights the difficulties in measuring fields in polar materials, where polarization and structure are intrinsically coupled. 

\begin{figure}
    \centering
    \includegraphics[width=\textwidth]{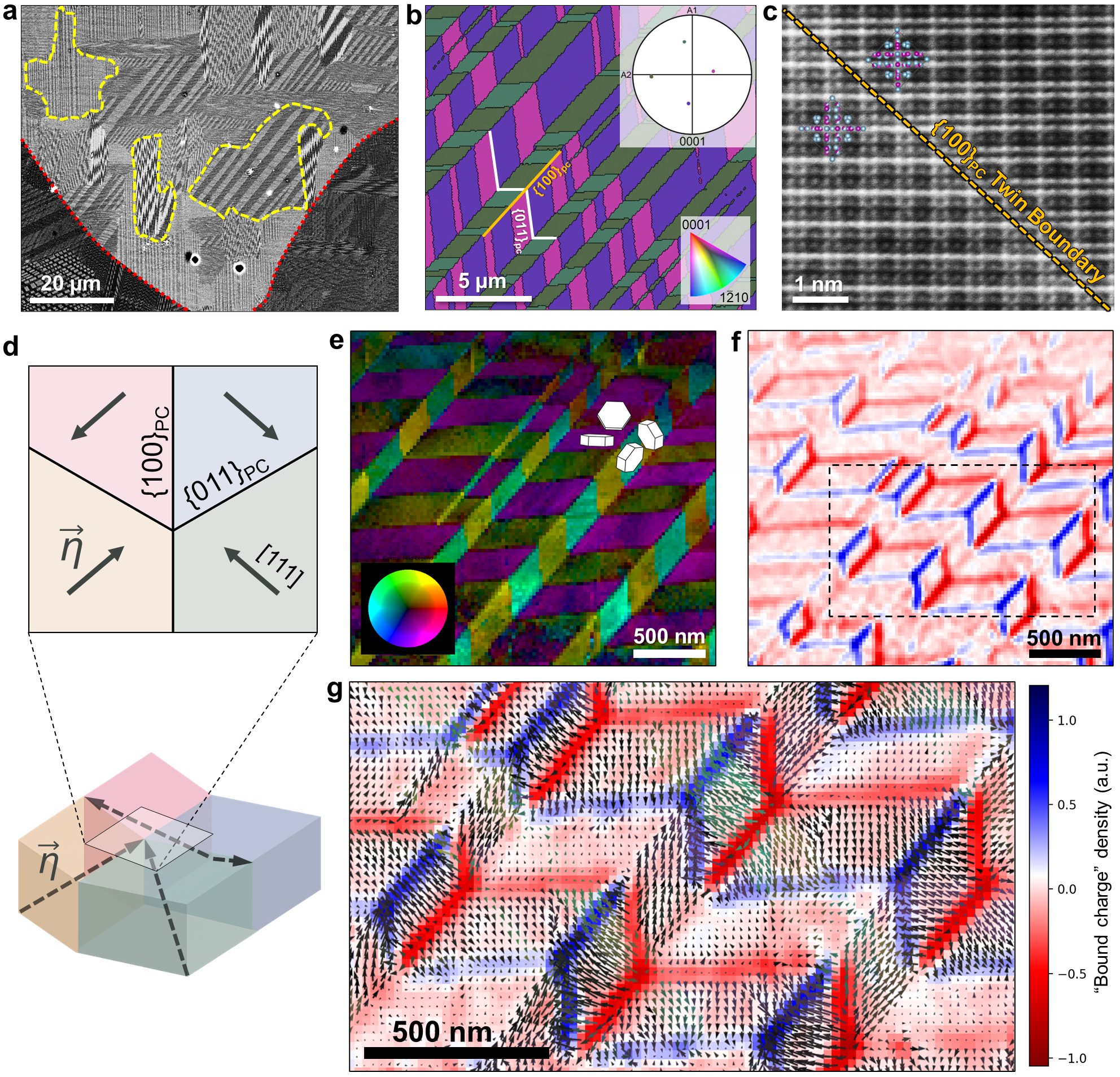}
    \caption{\textbf{Microstructure analysis and herringbone domain boundary characterization.} \textbf{a,} Backscattered electron (BSE) SEM micrograph of the as-cast Mn$_{51}$Al$_{49}$ microstructure, showcasing one large grain (grain boundary marked with red dotted line) and the diverse “colony” morphologies that make of up Rank-2 Laminate Twins. \textbf{b,} EBSD micrograph of characteristic herringbone domains, showing 4 different variants oriented $\sim90^{\circ}$ to each other with the domain boundaries marked with black line and the twin traces of $\{100\}_{PC}$ and $\{011\}_{PC}$ are indicated by orange and white lines, respectively; the corresponding pole figure is given in the inset. \textbf{c,} High-resolution HAADF-STEM micrograph taken of a $(100)_{PC}$-twin, viewed along the $[21\bar{3}\bar{2}]||[001]_{PC}$ zone axis. The twin boundary is marked with a dashed line as a guide to the eye, and polarization $\langle111\rangle$ axes given (going out of the plane) in light-blue arrows.}
    \label{fig:figure2}
\end{figure}
\begin{figure}[t]
    \contcaption{\textbf{d,} Schematic diagram depicting the mixed $\{100\}_{PC}/\{110\}_{PC}$ herringbone morphology. In \textbf{e,} DPC-STEM is utilized to view herringbone morphology. The vector field generated shows orientation domains, where each vector may be strongly influenced by diffraction contrast, due to structural changes between the variants. \textbf{f,} A ``bound charge” density map is made from displacement vectors in e by taking the divergence of the vector field. The blue/red domain boundaries correspond to negative/positive divergence values which are taken as head-to-head (H-H) and tail-to-tail (T-T) respectively. \textbf{g,} A superposition of displacement vector field and ``bound charge” density divergence map is given, where H-H/T-T domain wall configurations are visible for the blue/red domain boundaries.}
\end{figure}

Taking the divergence of the displacement-vector field in Fig. \ref{fig:figure2}e produces a ``bound charge” density map (see Fig. \ref{fig:figure2}f). Because the values are unitless and diffraction contrast generates the signal, the map should be interpreted only qualitatively. Notably, the blue/red boundaries in Fig. \ref{fig:figure2}f mark negative/positive divergence that correspond to H–H and T–T domain boundaries–configurations that are electrostatically non-neutral in dielectrics. Discontinuities in polarity lead to ``bound charge” which attract free carriers in a dielectric, though still virtually unexplored as a phenomenon in polar metals. The volume bound charge $\rho_{B}$ is calculated as $\rho_{B}=-\nabla \cdot \vec{P}$ where $\nabla$ is the divergence operator, and $\vec{P}$ in this instance is not expected to be electronic/ionic polarization but the measured deflection vector of the CoM due to diffraction-related intensity redistributions of the Bragg disk.  Nevertheless, H-H/T-T domain boundaries are clearly visible when the individual displacement vector lines are superimposed onto the ``bound charge” density map for a region of interest, as given in Fig. \ref{fig:figure2}g. It should be stated that the long-inclined twins are $\{100\}_{PC}$-type, which are cut longitudinally by $\{110\}_{PC}$-type domain boundaries: as such all $\{100\}_{PC}$ boundaries in this region should exhibit polar discontinuities, though some appear ‘neutral’ as they are only weakly edge-on to the lamella surface. It is important to note that the H-H and T-T domain boundaries form a distinct ‘interlocking Y pattern’ – a recurring motif referred to hereafter.

\subsection{Variable Surface Effects Determined by Domain Boundary Character}

The current picture of a polar metal suggests that any electric field induced by the electrical polarization (i.e. off-centering of ionic cores in a unit cell) is completely screened by free carriers which aggregate at domain boundaries and the sample surface \cite{Gu2023}. By this logic, even though Mn$_5$Al$_8$ is a common intermetallic, one could envision that the polar structure will influence the domain boundary characteristics. As such\cite{Gu2023}, there should be electron enrichment and depletion at H-H and T-T domain boundaries, respectively, while maintaining charge neutrality but comepnsating for the polar displacements of the ionic cores, which lead to localized changes in work function (as shown schematically in Fig. \ref{fig:figure3}a). Although charges are screened, the geometric polar displacements still induces local variations in “bound charge" which locally change the DOS. If that is the case, phenomena foreign to typical metallic materials may manifest on the surface commensurate to microstructures consisting of alternating H-H/T-T boundaries. The schematic in Fig. \ref{fig:figure3}b depicts the ‘interlocking Y pattern’ of   “charged”/ “charge-depleted” boundaries expected for herringbone morphologies, as observed by DPC-STEM Fig. \ref{fig:figure2}g. 

\begin{figure}
    \centering
    \includegraphics[width=\textwidth]{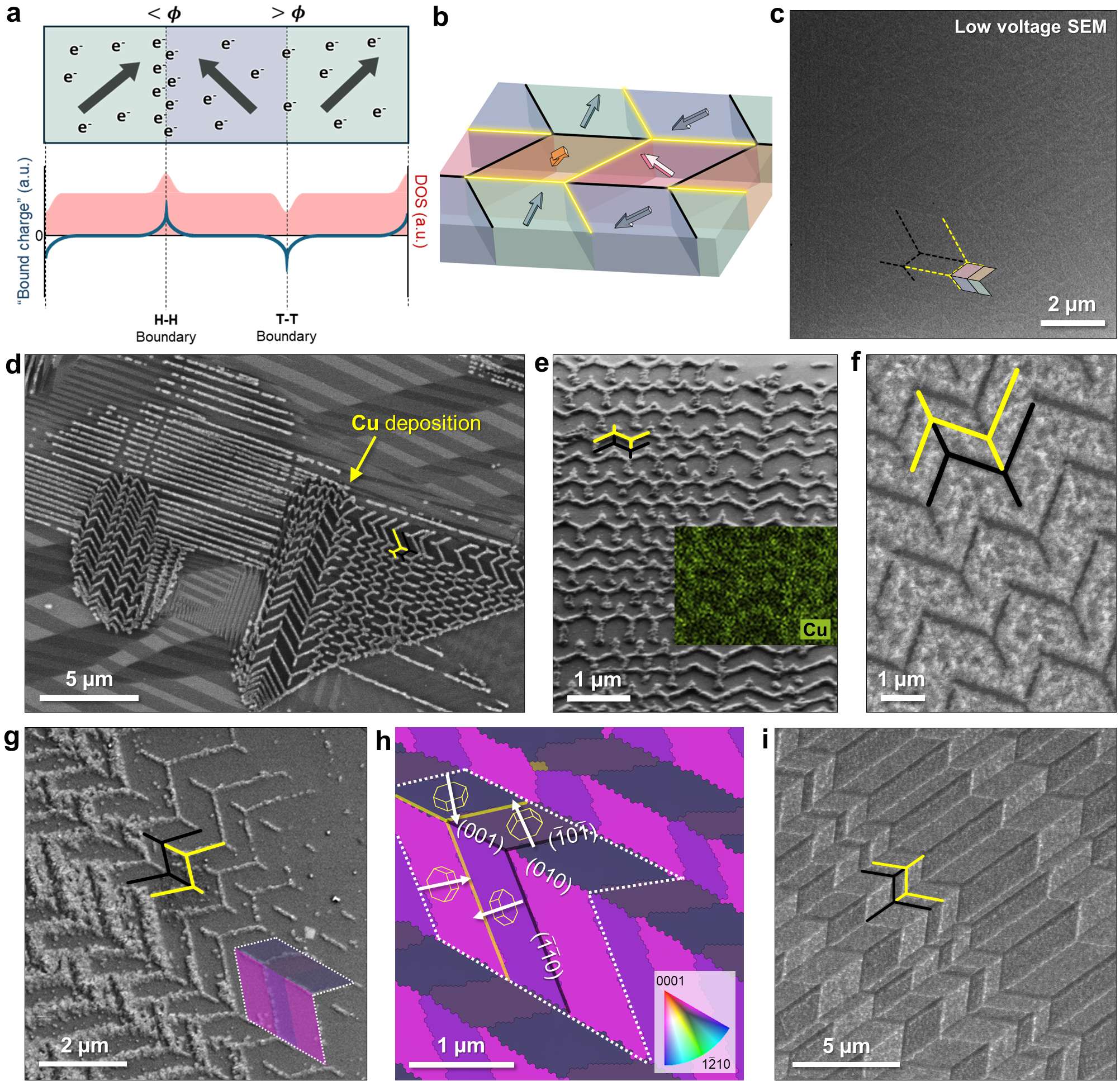}
    \caption{\textbf{Direct observation of surface effects at Mn$_5$Al$_8$ domain boundaries.} \textbf{a,} 2D schematic articulating metallic DPM domain boundaries, where surplus electrons (free charge carriers) can accumulate at H-H domain boundaries, commensurate with local “bound charge” deviations due to geometrical polarization. The variation in the DOS in turn yields a decrease of the work function for H-H domain boundaries, and the inverse is expected for T-T domain boundaries (increase in work function). \textbf{b,} A schematic illustration depicting the repeating ‘interlocking Y pattern’ outlining the herringbone morphology, with yellow/black lines corresponding to H-H/T-T domain boundaries, respectively \textbf{c,} SEM low voltage (0.5 kV) micrograph showing bright contrast at one set of (H-H) domain boundaries. \textbf{d,} BSE-SEM micrograph showing decoration of domain boundaries by deposition of copper particles. In this micrograph the herringbone morphology, and an adjacent morphology exhibiting linear “charged” boundaries (i.e. ‘backgammon’ domain morphology) is decorated with copper, though this patterning is not discussed herein (see Fig. S7c,d).}
    \label{fig:figure3}
\end{figure}
\begin{figure}[t]
  \contcaption{\textbf{e,} SEM in-lens micrograph at $60^{\circ}$ tilt showing the profile of the Cu-deposition ($
  \sim$100 nm width, $\sim$30 nm height) overlayed with a corresponding EDS Cu mapping inset. \textbf{f}, Herringbone region of a sample completely covered by copper deposition with a relief pattern abiding to one set of the domain boundaries (copper deposition avoided at regions with higher work function at T-T domain boundaries). \textbf{g,} SE-SEM micrograph showing Cu patterning along herringbone H-H domain boundaries. \textbf{h,} EBSD scan corresponding to a region with copper deposition observed. Trace analysis allows correct identification of $\{100\}/\{110\}$ planes, and twin domain orientation (where $[0001]_H||[111]_{PC}$ is the polarization axis). An inset of this region is overlaid onto the SEM micrograph of \textbf{g}, where superposition of $\langle111\rangle_{PC}$ shows the formation of H-H / T-T domain boundaries in agreement with copper deposition patterning. \textbf{i,} BSE-SEM micrograph showing strong variation between oxide build-up, commensurate with the herringbone ‘interlocking Y pattern’ suggesting variable surface reactivity between H-H/T-T domain boundary types. }% Continued caption
\end{figure}
The probing of charged ferroelectric domain walls with low-voltage SEM is well established \cite{Hunnestad2020}, where H-H charged domain walls exhibit stronger contrast than the bulk due to higher repulsion of incoming primary electrons (the inverse is true for electron-depleted T-T domain walls which should show up darker than the bulk). A low-voltage 0.5 kV SEM micrograph (see Methods) is given in Fig. \ref{fig:figure3}b, showing bright domain boundary contrast on half of the $\{100\}_{PC}/\{110\}_{PC}$ twins that make up the herringbone morphology, but no contrast within the domains. This observation suggests increased/decreased electron concentration commensurate with the ‘interlocking Y pattern’ domain morphology observed in Fig. \ref{fig:figure2}g. Note that the probed surface was polished by ion etching prior to the measurement to limit any topographical effects. The surface was characterized by X-ray photoelectron spectroscopy (XPS), where only a $<$$3$ nm-thick native oxide layer is observed (see Fig. S7a). Identical domain boundary contrast variation is visible also in electrostatic force microscopy (EFM) scan shown in Fig. S7b. The EFM scan in Fig. S7b herringbone region as observed with other techniques (SEM, EBSD \ref{fig:figure2}a,b). Furthermore, there is noticeable contrast variation observed across half of the $\{100\}_{PC}/\{110\}_{PC}$ domain boundaries (H-H and T-T), as commensurate with the ‘interlocking Y pattern’. EFM probing is sensitive to several factors \cite{Bellitto_2012,Lei2004}, and a robust understanding of contrast mechanism is complicated. Differences in contrast may be generated by orientation-dependent surface modifications, topography, active oxidation processes or irregular presence of passive oxidation layer. Another possibility is that intensity may be influenced by the local value of the work function: where an inverse relationship between electron concentration and work function is expected. 

Direct observation of (variable) surface reactivity along domain boundaries became evident by preferential copper deposition (Cu$^{2+}$ cations reducing onto the metal surface) along certain herringbone twins during metallographic preparation of the bulk MnAl samples (see Fig. \ref{fig:figure3}e-j). The copper deposition showed no preference for either $\{100\}_{PC}$ or $\{110\}_{PC}$ twins. Instead, copper deposition formed according to the “charged” ‘interlocking Y pattern’ as observed in previous measurements (see Fig. \ref{fig:figure2}g, \ref{fig:figure3}c,d). It is suggested that copper reduced by an electron transfer (redox) reaction deposited onto surfaces by galvanic displacement\cite{Djokic1996} with the lowest work function (i.e. regions with higher DOS just below the surface) \cite{searson1989}. A tilted-view ($60^{\circ}$) in-lens high-resolution SEM micrograph is provided in Fig. \ref{fig:figure3}f, coupled with an energy dispersive X-ray spectroscopy (EDS) elemental mapping overlay of copper. Further examples of the preferential copper deposition across various length scales are given in Fig. S8. Correspondingly, the ``charge-depleted” domain boundary pattern was also observed when abundant time for copper accumulation allowed near full copper coverage (see Fig. \ref{fig:figure3}g), where copper deposition formed the relief (inverse) of the ‘interlocking Y pattern’ (see Extended Data Fig. 8). Correlated EBSD analysis (see Fig. \ref{fig:figure3}g,h) confirmed the H-H/T-T preferential copper deposition, with the planes and polarity directions identified by a thorough trace analysis (shown in Fig. S10). This variable redox surface reactivity was also observed by preferential surface oxidation facilitation/suppression along respective domain boundaries, exhibiting the same ‘interlocking Y pattern’ (see Fig. \ref{fig:figure3}i and Fig. S9d). The same logic applies herein, where lower work function would facilitate oxide growth while electron-depleted regions would inhibit oxidation.

\subsection{Chemical Space of Metallic DPMs}
\begin{figure}[H]
    \centering
    \includegraphics[width=\textwidth]{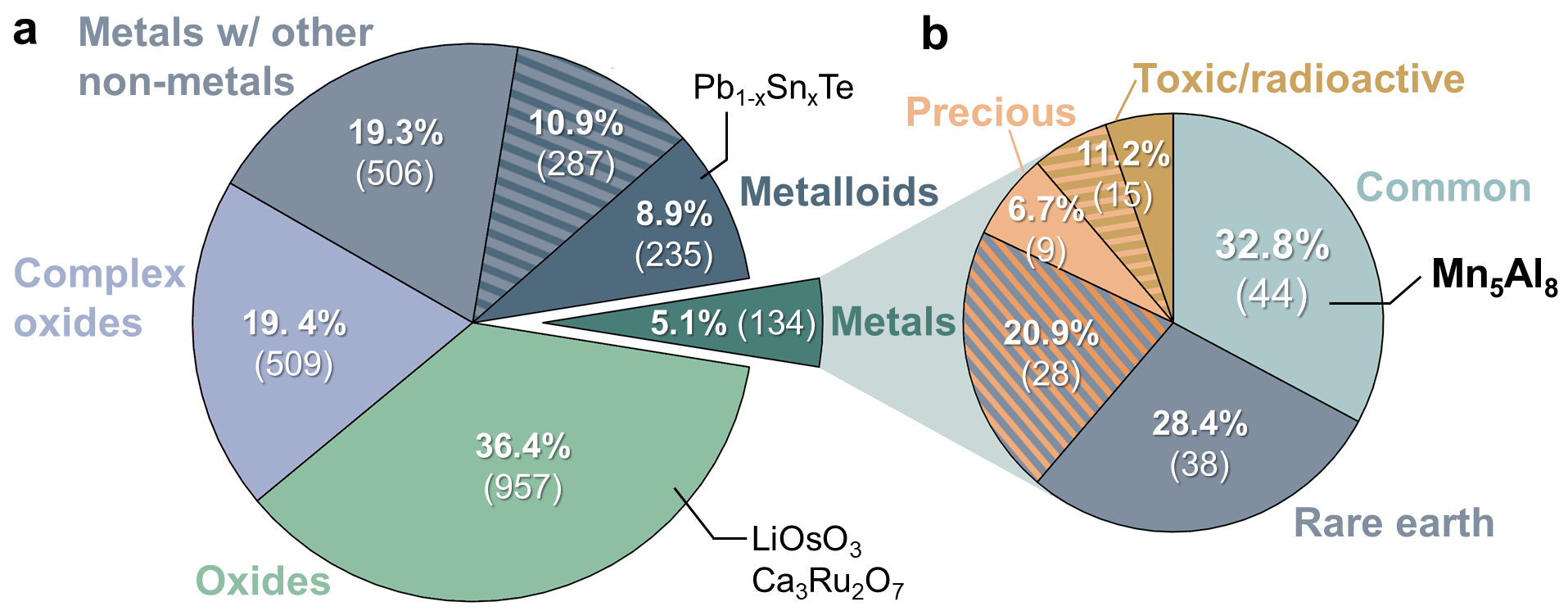}
    \caption{\textbf{Chemical sorting of polar compounds and metallic polar metals.} \textbf{a,} A chemical sorting of all compounds in one of the ten polar crystal classes (point groups 1, 2, $m$, $mm2$, 3, $3m$, 4, $4mm$, 6, $6mm$) in the ICSD ($\sim$2600 unique compounds): compound containing oxygen are listed as oxides, whereas complex oxides here include any oxyanion-species containing oxide group (e.g., carbonates, sulfates). Compounds containing other non-metal species (e.g. carbides, nitrides, sulfides) are listed as such. Compounds listed in the metalloid category contain one of these six species (B, Si, Ge, As, Sb, Te,), and metallic compounds comprise only metallic alkaline, alkaline earth, transition, post-transition, and lanthanide species (5.1\% or $\sim$134 in total). Using the common notion of chemical elements as metals, metalloids, and nonmetals provides a simple, but useful classification of the predominant bonding type by chemical composition. \textbf{b,} Further sorting of metallic compounds into subcategories of rare earth, precious, toxic/radioactive, and common (of which Mn$_5$Al$_8$ is one). Hatched regions correspond to compounds containing species in both adjacent groups.}
    \label{fig:figure4}
\end{figure}
\begin{table}[h]
\caption{\label{tbl:1} \textbf{Notable Examples of Metallic Distortive Polar Metals.} Metallic compounds identified through a chemical sorting of polar space groups in the Inorganic Crystal Structure Database (ICSD). All compounds undergo a transition from centrosymmetry to a polar structure at temperature T$_C$, with both parent/product space groups listed as well. M-monoclinic, O-orthorhombic, H-hexagonal, \textdagger-metastable, *-peritectoid. }
%\fontsize{9.5pt}{50pt}
\centering
\begin{tabular}{cccccc}
\hline
\textbf{Compound} & \textbf{Lattice} & \textbf{$T_C$} & \makecell{\textbf{Parent space} \\ \textbf{group}} & \makecell{\textbf{Product space} \\ \textbf{group}} & Ref. \\
\hline
Mn$_5$Al$_8$ (Mn$_{51}$Al$_{49}$) & H & $\sim880^{\circ}$C & $I\bar{4}3m$ & $R3m$ & - \\
Ce$_{0.98}$Pt$_2$Sn$_{1.65}$ & M & $\sim250^{\circ\dagger}$C & $P4/mmm$ & $P2_1$ & \cite{LIU2000} \\
$\beta$-Ce$_{3}$Al$_{11}$ & O & $\sim1006^{\circ}$C & $I4/mmm$ & $Immm$ & \cite{Gschneidner1988} \\
Au$_{0.939}$Cu$_{1.25}$Zn$_{1.81}$ & O & $\sim25^{\circ}$C & $Fm3m$ & $P2mm$ & \cite{DUGGIN1966} \\
Cu$_{1.45}$Ni$_{0.05}$Al$_{1.5}$ & O & $\sim25^{\circ}$C & $Fm3m$ & $P2mm$ & \cite{DUGGIN1966} \\
$\beta$-BiK$_3$ & H & $277^{\circ}$C & $Fm\bar{3}m$ & $P6_3cm$ & \cite{kerber1998} \\
$\beta'$-Mg$_2$Al$_3$ & H & $220^{\circ}$C & $Fd\bar{3}m$ & $R3m$ & \cite{Steurer2007} \\
$\beta_1$-InMg$_3$ & H & $337^{\circ}$C & $Fm3m + Pm3m$ & $R3m$ & \cite{NayebHashemi1985} \\
NiZn$_3$ & O & $\sim525^{\circ}$C & $I\bar{4}3m$ & $Acm2$ & \cite{NOVER1980} \\
$\zeta_2$-Al$_3$Cu$_4$ & O & $575^{\circ}$C & $Fmm2$ & $Imm2$ & \cite{Gulay2003} \\
m-Co$_4$Al$_{13}$ & M & $1083^{\circ}$C$^*$ & $C2/m + Cmcm$ & $Cm$ & \cite{goedecke1996} \\
Au$_5$Zn$_3$ & O & $500^{\circ}$C-$300^{\circ}$C$^{\dagger}$ & $Pm\bar{3}m$ & $Pmc2_1$ & \cite{iwasaki1965} \\
\hline
\end{tabular}
\end{table}

The unexpected domain boundary behavior spurs further interest to identify similar DPM compounds. It is worth mentioning that metals in polar crystal classes, historically, were mainly studied for superconducting/transport properties \cite{Steurer2007,Zaremba2004,lawson1978,Bauer2004}, due to their lack of an inversion center \cite{Yip2014,bulaevskii1976} but not for their polar domain state or domain boundaries. Therefore, to develop and contextualize the chemical space of metallic polar metals, we perform a chemical sorting of polar compounds in the Inorganic Crystal Structure Database (ICSD) (see Fig. \ref{fig:figure4}a). Wherein, compounds comprising strictly metallic species in one of the polar space groups, are by rudimentary approximation polar metals. Although representing a relatively narrow fraction ($5.1\%$) of the total polar compound list, at least 134 unique metallic polar metals were identified in this sorting, which were sorted based on further classifications (see Fig. \ref{fig:figure4}b). Of those, approximately 20$\%$ are DPMs, as they exhibit a symmetry-lifting transformation into a polar crystal class. Notable examples of metallic DPMs are shown in Table \ref{tbl:1} with transition temperature ($T_C$) and parent/product space groups listed. Several compounds stick out, such as Au$_{0.939}$Cu$_{1.25}$Zn$_{1.81}$ \cite{DUGGIN1966}, Cu$_{2.9}$AlNi$_{0.1}$ \cite{DUGGIN1966}, which were classically studied martensitic transformations. The full sorting of metallic DPMs found is provided as supplementary information. Of course, it should be restated that DPMs can exist in non-metallic chemistries and are not discussed in the context of this study, for example the topological crystalline insulator Pb$_{1-x}$Sn$_x$Te \cite{Shi2024}, oxides like LiOsO$_3$ \cite{Shi2013}, Ca$_3$Ru$_2$O$_7$ \cite{Lei2018}, and the multiferroic-like Pb$_2$CoOsO$_6$ \cite{Princep2020}, although metallic electrical conductivity in non-metals is rare \cite{Benedek2016}.

\section{Discussion}\label{sec:discussion}

The empirical evidence uncovered in this study points to the concept of ``bound charge” influencing carrier concentration at domain boundaries in the distortive polar metal Mn$_5$Al$_8$. The concept of bound charge is well-established in insulating ferroelectrics with strong de-polarizing electrostatic fields, where electronic band-bending can create an excess of (conduction-band) electrons and (valence-band) holes at H-H and T-T boundaries, respectively \cite{iguez2020}. Since (polar) metals exhibit surplus free-carriers and disqualify electron band-bending as a mechanism, the accumulation or depletion of electrons at “charged” boundaries is unintuitive. Nevertheless, the influence of geometric polarization discontinuities on free carriers is non-negligible, as subsurface variations in the DOS adequately describes the observed copper decoration and surface oxidation behavior along H-H (while absent at T-T) boundaries. Preferential domain boundary surface decoration in ferroelectrics is known and can occur for several reasons including electric field gradient interaction with charged particles \cite{Hanson2006} and charge-induced reduction or oxidation \cite{Burbure2010}. Since no electrostatic field may be emitted from a metal, the copper deposition is expected to have formed by electron transfer reduction on more reactive surface regions with a lower electronic work function \cite{Lin2023,Chen2024}. Given the standard potential of copper reduction and metal work function values\cite{StandardPotentials}, we postulate H-H/T-T boundaries induce a variation of $\pm$0.5 eV.

Efforts were made to explore all potential influences on the observed surface effects at domain boundaries. For example, effects related to magnetism were ruled out, as hysteresis measurements displayed a characteristically paramagnetic response (see Fig. S1d). The degree of chemical ordering was not characterized in this study, but nevertheless may have an effect on the observed domain boundary behavior. The influence of strain could be more involved, since electronic work functions may be lowered at grain boundaries \cite{Khisamov2013,Orlova2018}, due to lattice distortions and grain boundary defects (e.g. dislocations). However, the examined $\{100\}_{PC}/\{110\}_{PC}$ herringbone microstructure exhibited only the necessary localized elastic strain at all boundaries without noticeable differences pertaining to their geometric polarization discontinuity (see Fig. S3). Further, the prevailing ‘interlocking Y pattern’ observed for the herringbone morphology by 4D-STEM DPC (Figure \ref{fig:figure2}), preferential copper decoration, and surface oxidation (Figure \ref{fig:figure3}) was not commensurate with elastic strain. In addition, work function differences between adjacent domain surfaces may lead to Fermi level equilibration at boundaries, forming a “contact potential” at twin interfaces \cite{hutchinson1968}. However, this alone cannot describe the bi-modal electronic behavior observed between H-H and T-T boundaries, where subtle shifts in crystallography associated with polar axis orientations between domains may aide in local variation in the DOS. Recent efforts to study electronic states of twin and grain boundaries in Cu \cite{ni2024} showed surplus carriers at twin boundaries vis-à-vis bulk, but found no complementary depleted twin boundaries. Thus, while the patterning of deposited Cu and oxidation is commensurate with the H-H or T-T ``charged”/``charge-depleted” boundaries it is still mechanistically unclear, and geometric polarization discontinuities may affect free carrier density. The role of complex electronic structure stabilizing $\gamma$,$\gamma_{2}$ Hume-Rothery\cite{Pettifor1997} phases may also influence domain boundary behavior. 

The symmetry-lifting transformation in Mn$_5$Al$_8$ and other metallic DPMs is worthy of future study, since DPMs in fully metallic compounds may challenge the ‘weak-coupling principle’ \cite{hickoxyoung2021}, where sublattice de-coupling is unexpected since metallic bonding suggests all species contribute free carriers at the Fermi level $E_F$,. To date, metallic oxides have been the focus of DPM studies due to the mechanistic de-coupling allowed in perovskite or Ruddlesden-Popper sublattices \cite{Kim2016}. Despite maintaining metallic conductivity (i.e. single peak of optical conductivity as frequency $\omega\rightarrow0$) \cite{hockoxyoung2023}, oxides are oftentimes ionically bonded (limiting free carriers, making polar metallic oxides somewhat rare \cite{Benedek2016,Puggioni2014}). For example, in LiOsO$_3$ it is primarily the under-bonded A-site Li atoms that undergo polar distortion along the trigonal $[001]$ axis \cite{Shi2013}, whereas metallicity is mainly derived from potentially correlated O and Os orbitals \cite{Laurita2019,Shi2013,LoVecchio2016}. This is similar to the Ca ions that rotate to induce polarization in Ca$_3$Ru$_2$O$_7$ but contribute few states to the Fermi level $E_F$ \cite{lei2017}. In addition, since the symmetry-reducing transformation in Mn$_5$Al$_8$ is ferroelastic by space group considerations \cite{Hlinka2016}, applied pressure can act as a potential avenue to allow polar domain flipping as is currently being explored in other polar metals \cite{Zabalo2021,peng2024}. Moreover, unlike oxides, metallic polar metals are amenable to traditional metallurgical processing, including facile synthesis, alloying, and thermomechanical treatments. The identification of various DPMs (Table \ref{tbl:1}) with extensive chemical space (Figure \ref{fig:figure4}) and possible functionalizable nature of domain boundaries in metallic DPMs adds an unexplored but meaningful dimension to metals that combine a polar crystal structure in a sea of electrons.

\section*{4 Conclusion}\label{sec:conclusion}
Distortive polar metals undergo a symmetry-lifting transformation into a polar space group and in the process form technologically-relevant domains and domain boundaries. The intermetallic DPM Mn5Al8 ($R3m$ space group) exhibits $\{$100$\}$/$\{$110$\}$ twins in both “charged” H-H and T-T domain boundary configurations, with polar axes aligned along [111] of the pseudo-cubic unit cell. Advanced electron microscopy techniques (4D-STEM DPC and low-voltage SEM) allow visualization of these boundaries, and reveal an ‘interlocking Y pattern’ for H-H/T-T configurations. Here, we find that domain boundaries exhibit variable surface reactivity based on their character, as revealed through preferential galvanic deposition of copper (as well as surface oxidation) along certain boundaries and avoidance along others. This behavior can be explained by an increase and depletion of electronic density of states for H-H/T-T boundaries respectively, effectively lowering or raising the electronic work function. These findings reveal unexpected and potentially functionalizable domain boundaries in seemingly “simple” metallic materials. A sorting of polar compounds by chemistry allows further identification of DPMs, effectively broadening the chemical space of this polar metal subclass. Overall, the observed phenomena prompt a broader discussion on the mechanisms governing symmetry-lifting transitions into a polar structure (e.g. weak-lattice hypothesis) and the effects of polar discontinuities in metals.

\section*{5 Methods}\label{sec:methods}
\subsection*{Material synthesis}
In this study, stoichiometric and off-stoichiometric Mn$_5$Al$_8$ alloys (Mn$_{50}$Al$_{50}$) were induction-melted under protective argon atmosphere at $\sim1350^{\circ}$C and cast into copper rod-shaped crucibles. Al pellets (99.7 wt\% pure) and Mn flakes (99.8 wt\% pure, with $\sim$0.15 wt\% Fe, Si, Se impurities) were used as the raw materials. An excess of 2 wt.\% of Mn was added to the melt, as the vapor pressure of Mn is relatively high, and some evaporates during the melting stage of the synthesis. 

\subsection*{XRD}
X-ray diffraction was performed on Mn$_{51}$Al$_{49}$ samples using a Rigaku Smartlab 9kW diffractometer utilizing a DHS1100 Anton Paar heated stage sample holder for room temperature ($30^{\circ}$C) and elevated temperature experiments (600$^{\circ}$-1000$^{\circ}$C). The sample was held in an inert atmosphere (helium), with heating/cooling rate set to 20 K/min. A Cu X-ray source, utilizing Bragg-Brentano geometry probing 20-100$^{\circ}$ $2\theta$, with scan speed 6 $^{\circ}$/min and 0.01$^{\circ}$ step size. Samples were mirror polished prior to diffraction experiments.

\subsection*{Metallographic preparation and copper deposition}
MnAl samples were mounted in conductive Bakelite and ground and polished using SiC abrasive papers from 220 up to 4000 grits, followed by diamond slurry polishing (with 3 and 1 µm) and final mechano-chemical polishing by silica nanoparticles (0.05 µm) suspension (OP-S). In some instances, copper deposition was introduced during OP-S polishing on contaminated pads (yielding Cu2+ in the solution). This was carried out in two ways: either with limited deposition particularly on ``charged” domain boundaries by a short 3 min polishing duration or nearly full surface coverage by polishing for an extended duration of 20 min (with the deposition avoiding ``charge-depleted” regions). Other selected samples were polished further by a final step of Ar ion etching using Gatan PECS at a grazing angle of 3° with 2 kV and 40 mA operating conditions.

\subsection*{ICP-OES}
Inductively-coupled Optical emission spectroscopy was performed to verify chemical composition of bulk melts. Utilizing an iCAP PRO XP Duo from Thermo Fisher Scientific, measurements were carried out on as-cast samples (Mn$_{51}$Al$_{49}$) to verify accurate chemical composition, and that significant macroscale chemical segregation had not occurred. 

\subsection*{SEM and EDS}
The microstructure of the metallographically prepared samples was examined by a Zeiss Merlin thermal field emission high-resolution scanning electron microscope, operated at 10 kV and 2.2 nA for BSE mode and 2 kV 2.2 nA for the SE and in-lens modes. In addition, the effects of acceleration voltage (5-20 kV) and tilting by $\pm 2^{\circ}$ on the domain structure appearance were analyzed. EDS measurements (using an Oxford instruments EDX detector) were performed at 10 kV to assess the overall chemical composition of the Mn$_5$Al$_8$ and Mn$_{51}$Al$_{49}$ metallographically prepared cast alloys and at 2 kV with a 60 $^{\circ}$ tilt when mapping copper deposited on the domain boundaries. Low voltage imaging in SEM (ZEISS GeminiSEM 450) of Ar ion-polished surfaces was conducted using the in-lens detector operated at 0.5 kV, 100 pA with a working distance of 2 mm. 

\subsection*{EBSD}
Electron backscatter diffraction analysis of the Mn$_5$Al$_8$ domain structure was carried out in the same Zeiss Merlin high-resolution SEM. The EBSD maps were collected with a step (pixel) size of 50 nm using an EDAX DigiView 5 camera. As the 4 different $\{111\}_{\gamma}||\{0001\}_{\gamma_2}$ twin variants that form from one $\gamma$ grain are very similar in plane position (only $\sim$1 \% of strain along $[111]_R$) indexing of the patterns is much more reliable when taking into account the brightness and fine structure of the Kikuchi bands. Therefore, spherical indexing \cite{LENTHE2019} was employed instead of a Hough-transform based approach to reindex the Kikuchi patterns. Nevertheless, despite the use of spherical indexing with the highest number of spherical harmonics (bandwidth 255), the software was not able to distinguish the negative and positive polar direction of the crystal structure. Therefore, the crystal orientations were analyzed as if the structure would contain a center of symmetry.

\subsection*{TKD}
Transmission Kikuchi diffraction measurements were performed on a lamella prepared by FIB, containing the herringbone structure (from the same grain region used for the TEM analysis), using a Zeiss Merlin high-resolution microscope equipped with a VelocityGen1 camera. The Kikuchi patterns were acquired employing an acceleration voltage of 30 kV and scanning with a step size of 10 nm. The Kikuchi patterns data sets were reindexed by spherical indexing as mentioned above. 

\subsection*{PPMS}
Physical property measurement system (PPMS) resistivity measurements were performed with the four probe method on polished surface of samples, cut into rectangularly shaped plates with dimensions of $5\times5\times1$ mm$^3$. Platinum wires were employed to establish electrical contact to the sample. To ensure good ohmic contact silver epoxy was used to attach the wires. Magnetic measurements, were done using PPMS 14 (Quantum design) with the vibrating sample magnetometer (VSM) option. 

\subsection*{DFT}
The density functional theory (DFT), spin polarized calculations on Mn$_5$Al$_8$ hexagonal cell containing 78 atoms, were performed in S/PHI/nX library \cite{BOECK2011}. For the calculations, we use the projector augmented wave pseudopotentials taken from the VASP library and the Perdew-Burke-Ernzerhof (PBE) \cite{perdew1996} exchange-correlation functional. We ensured careful convergence with respect to the k points and the energy cutoff for plane waves to yield an accuracy of $10^{-3}$ eV in the relevant energies. A k-point sampling of $6\times6\times6$ and a plane-wave energy cutoff of 320 eV was used to get a converged electronic density. For band structure calculation we used the full path along the high symmetry points $\Gamma$, L, T, P$_0$, P$_2$, $\Gamma$ and F, with 300 k-points. Using the band structure, the density of states (DOS) was calculated using a gaussian smearing parameter of 0.1 eV.

\subsection*{EFM}
Electrostatic force microscopy (EFM) was operated in a dual-pass measurement approach, where the resulting changes in the amplitude and phase of the cantilever at the oscillation frequency are recorded in the EFM amplitude and phase channels. A constant height was maintained to minimize the effects of other interactions, such as van der Waals forces. A conductive AC silicon probe (SPARK 70 Pt) was used for the measurements with an applied voltage of 1 - 2 V and an oscillation frequency of 0.5 Hz - 5Hz. The measurements were carried out with an Oxford Instruments Cypher Environmental System, in a dry N$_{2}$ gas atmosphere.

\subsection*{XPS}
X-ray photoelectron spectroscopy (XPS) of clean and copper patterned surfaces was performed in a Physical Electronics PHI Quantera II spectrometer. The spectrometer is equipped with an Al K$\alpha$ source at 1486.6 eV. The survey measurement was performed with a launch angle of 45$^{\circ}$. Peak fitting of core spectra from XPS studies was performed using CasaXPS software.

\subsection*{Lamellae preparation by FIB}
The TEM lamella preparation was carried out on a Dual Beam FIB/SEM Scios2 from Thermo Fisher Scientific. A cross section lamella was taken from the region of interest from the previously metallographically prepared specimen. The lamella thickness was reduced step by step with ever decreasing currents at an acceleration voltage of 30kV. In order to obtain sufficient electron transparency without influencing the material too much by the ion bombardment, the lamellae were finally thinned with 5kV and 48pA.

\subsection*{TEM}
The conventional TEM investigations were carried out on an image corrected Titan Themis 80-300 and the STEM investigations on a probe corrected Titan Themis 60-300 (Thermo Fisher Scientific) with an acceleration voltage of 300kV. Both microscopes are equipped with a high-brightness field emission gun and the probe corrected additionally with a gun monochromator. The STEM images were recorded at a probe current of about 70 pA with a high-angle annular dark field (HAADF) detector (Fishione Instruments Model 3000). The collection angles for the HAADF images were set to 78–200 mrad using a semi-convergence angle of 23.8 mrad.

\subsection*{Four-dimensional STEM (4DSTEM) data collection}
The 4D-STEM data were also acquired in a probe corrected Titan microscope at 300 kV. We collected the entire convergent beam electron diffraction (CBED) pattern as a two-dimensional (2D) image for each probe position during scanning. These images were taken using an electron-microscope pixel-array detector (EMPAD) with a readout speed of 0.86 ms per frame and a linear electron response of 1,000,000:1. Each CBED image has a size of 128$\times$128 pixels$^2$. All data sets were acquired with a semi-convergence angle of 23.6 mrad, a defocus value of approximately 0 nm, and a camera length of 300 mm. The exposure time was 1 ms per frame. Beam scanning was synchronized with the EMPAD camera with a scanning step size of 18 pm and a field of view of 2.3$\times$2.3 nm$^2$. We calibrated the CBED pattern in reciprocal space using standard Au nanoparticles. Here, each pixel in the CBED pattern is 2.0 mrad.

\subsection*{4D-STEM DPC analysis}
We probe the local electric fields using 4D-STEM DPC microscopy, an imaging technique that measures the relative electron probe shifts observed on CBED patterns caused by local electric and magnetic fields \cite{hachtel2018,MULLERCASPARY2017,CHAPMAN1978,gao2019,mullercaspary2014,shibata2012}. When an electron probe traverses an electric field, it is deflected by the Lorentz force on the negatively charged electrons. By measuring the CoM shift of the transmitted probe in diffraction space, we can calculate the corresponding change in momentum. With appropriate modeling, the underlying electric field in the material can then be derived. According to the Ehrenfest theorem \cite{ehrenfest1927}, which also holds in quantum mechanics \cite{MULLERCASPARY2017}, the momentum gained by the electron probe is opposite in direction and proportional in magnitude to the local electric field. Gauss’s law links the divergence of electric field to the local charge density. The weak-phase-object approximation \cite{pennycook2011} – which assumes that electrons pass through a thin specimen without changing their velocity along $z$ – is not fully satisfied for our lamella. Consequently, we present the 4D-STEM DPC data as qualitative maps rather than absolute, quantitatively calibrated electric-field or charge-density distributions. 

\section*{Acknowledgements}
The authors would like to thank Ray McQuaid and Conor McCluskey, from Queens University of Belfast for insightful discussions on domain boundaries. In addition, the authors would like to thank Pavel Volkov from the University of Connecticut for fruitful discussions about polar metals. The authors are also grateful for insightful exchanges with Andreas Leineweber. A.S. gratefully acknowledges the support of a Max-Planck-Gesellschaft Scholarship. B.R. gratefully acknowledges the support of a Minerva Stiftung Fellowship and Alexander von Humboldt Foundation Fellowship. The Authors would like to thank Rebecca Renz for help with the schematic illustrations. Fernando Maccari provided critical help with chemical analysis among other techniques and discussion, which the authors are grateful for. Adam Miles provided insight into the copper deposition. The authors extend thanks to several members of the Max Planck Institute for Sustainable Materials for technical assistance: Dennis Klapproth for synthesizing the MnAl casts, Katja Angenendt for performing the TKD measurements, Christian Broß and Zakarya Meshou for ion polishing, and Benjamin Breitbach for X-ray diffractometry experiments.

\section*{Author contributions}
 A.S. and B.R conceptualized the study. B.R. prepared the bulk samples. B.R., and A.H. performed metallographic preparation and copper deposition. A.S. analyzed the XRD data. I.R. performed the PPMS measurements. B.R., S.Z., M.J.K., and M.R. performed SEM and EBSD investigation and analysis. S.Z. and B.R. performed the low voltage SEM examination. A.S., S.Z., and M.R. conducted the crystallographic analysis. P.W. prepared the TEM samples. A.S., X.Z., and P.W. performed TEM examination and analysis. X.Z. performed 4D-STEM measurements and was assisted by A.S. with data analysis. P.J.K. performed EFM and XPS analysis. A.V. supported the EFM investigation. S.K. performed the DFT calculations. A.S. drafted the manuscript, and took care of major revisions with the help of B.R. All authors discussed the results and contributed to the writing, revision and editing of the submitted manuscript.

\bibliography{sn-bibliography}% common bib file


\begin{thebibliography}{107}
\ifx \bisbn   \undefined \def \bisbn  #1{ISBN #1}\fi
\ifx \binits  \undefined \def \binits#1{#1}\fi
\ifx \bauthor  \undefined \def \bauthor#1{#1}\fi
\ifx \batitle  \undefined \def \batitle#1{#1}\fi
\ifx \bjtitle  \undefined \def \bjtitle#1{#1}\fi
\ifx \bvolume  \undefined \def \bvolume#1{\textbf{#1}}\fi
\ifx \byear  \undefined \def \byear#1{#1}\fi
\ifx \bissue  \undefined \def \bissue#1{#1}\fi
\ifx \bfpage  \undefined \def \bfpage#1{#1}\fi
\ifx \blpage  \undefined \def \blpage #1{#1}\fi
\ifx \burl  \undefined \def \burl#1{\textsf{#1}}\fi
\ifx \doiurl  \undefined \def \doiurl#1{\url{https://doi.org/#1}}\fi
\ifx \betal  \undefined \def \betal{\textit{et al.}}\fi
\ifx \binstitute  \undefined \def \binstitute#1{#1}\fi
\ifx \binstitutionaled  \undefined \def \binstitutionaled#1{#1}\fi
\ifx \bctitle  \undefined \def \bctitle#1{#1}\fi
\ifx \beditor  \undefined \def \beditor#1{#1}\fi
\ifx \bpublisher  \undefined \def \bpublisher#1{#1}\fi
\ifx \bbtitle  \undefined \def \bbtitle#1{#1}\fi
\ifx \bedition  \undefined \def \bedition#1{#1}\fi
\ifx \bseriesno  \undefined \def \bseriesno#1{#1}\fi
\ifx \blocation  \undefined \def \blocation#1{#1}\fi
\ifx \bsertitle  \undefined \def \bsertitle#1{#1}\fi
\ifx \bsnm \undefined \def \bsnm#1{#1}\fi
\ifx \bsuffix \undefined \def \bsuffix#1{#1}\fi
\ifx \bparticle \undefined \def \bparticle#1{#1}\fi
\ifx \barticle \undefined \def \barticle#1{#1}\fi
\bibcommenthead
\ifx \bconfdate \undefined \def \bconfdate #1{#1}\fi
\ifx \botherref \undefined \def \botherref #1{#1}\fi
\ifx \url \undefined \def \url#1{\textsf{#1}}\fi
\ifx \bchapter \undefined \def \bchapter#1{#1}\fi
\ifx \bbook \undefined \def \bbook#1{#1}\fi
\ifx \bcomment \undefined \def \bcomment#1{#1}\fi
\ifx \oauthor \undefined \def \oauthor#1{#1}\fi
\ifx \citeauthoryear \undefined \def \citeauthoryear#1{#1}\fi
\ifx \endbibitem  \undefined \def \endbibitem {}\fi
\ifx \bconflocation  \undefined \def \bconflocation#1{#1}\fi
\ifx \arxivurl  \undefined \def \arxivurl#1{\textsf{#1}}\fi
\csname PreBibitemsHook\endcsname

%%% 1
\bibitem[\protect\citeauthoryear{(Editor)~Hanh}{2005}]{ITC}
\begin{botherref}
\oauthor{\bsnm{(Editor)~Hanh}, \binits{T.}}:
In: International Tables of Crystallography Volume A.
Springer
(2005)
\end{botherref}
\endbibitem

%%% 2
\bibitem[\protect\citeauthoryear{Ok et~al.}{2006}]{Ok2006}
\begin{barticle}
\bauthor{\bsnm{Ok}, \binits{K.M.}},
\bauthor{\bsnm{Chi}, \binits{E.O.}},
\bauthor{\bsnm{Halasyamani}, \binits{P.S.}}:
\batitle{Bulk characterization methods for non-centrosymmetric materials: second-harmonic generation{,} piezoelectricity{,} pyroelectricity{,} and ferroelectricity}.
\bjtitle{Chemical Society Reviews}
\bvolume{35},
\bfpage{710}--\blpage{717}
(\byear{2006})
\doiurl{10.1039/B511119F}
\end{barticle}
\endbibitem

%%% 3
\bibitem[\protect\citeauthoryear{Bhowal and Spaldin}{2023}]{bhowal2023}
\begin{barticle}
\bauthor{\bsnm{Bhowal}, \binits{S.}},
\bauthor{\bsnm{Spaldin}, \binits{N.A.}}:
\batitle{Polar metals: Principles and prospects}.
\bjtitle{Annual Review of Materials Research}
\bvolume{53}(\bissue{Volume 53, 2023}),
\bfpage{53}--\blpage{79}
(\byear{2023})
\doiurl{10.1146/annurev-matsci-080921-105501}
\end{barticle}
\endbibitem

%%% 4
\bibitem[\protect\citeauthoryear{Anderson and Blount}{1965}]{Anderson1965}
\begin{barticle}
\bauthor{\bsnm{Anderson}, \binits{P.W.}},
\bauthor{\bsnm{Blount}, \binits{E.I.}}:
\batitle{Symmetry considerations on martensitic transformations: ``ferroelectric" metals?}
\bjtitle{Physical Review Letters}
\bvolume{14},
\bfpage{217}--\blpage{219}
(\byear{1965})
\doiurl{10.1103/PhysRevLett.14.217}
\end{barticle}
\endbibitem

%%% 5
\bibitem[\protect\citeauthoryear{Shi et~al.}{2013}]{Shi2013}
\begin{barticle}
\bauthor{\bsnm{Shi}, \binits{Y.}},
\bauthor{\bsnm{Guo}, \binits{Y.}},
\bauthor{\bsnm{Wang}, \binits{X.}},
\bauthor{\bsnm{Princep}, \binits{A.J.}},
\bauthor{\bsnm{Khalyavin}, \binits{D.}},
\bauthor{\bsnm{Manuel}, \binits{P.}},
\bauthor{\bsnm{Michiue}, \binits{Y.}},
\bauthor{\bsnm{Sato}, \binits{A.}},
\bauthor{\bsnm{Tsuda}, \binits{K.}},
\bauthor{\bsnm{Yu}, \binits{S.}},
\bauthor{\bsnm{Arai}, \binits{M.}},
\bauthor{\bsnm{Shirako}, \binits{Y.}},
\bauthor{\bsnm{Akaogi}, \binits{M.}},
\bauthor{\bsnm{Wang}, \binits{N.}},
\bauthor{\bsnm{Yamaura}, \binits{K.}},
\bauthor{\bsnm{Boothroyd}, \binits{A.T.}}:
\batitle{A ferroelectric-like structural transition in a metal}.
\bjtitle{Nature Materials}
\bvolume{12}(\bissue{11}),
\bfpage{1024}--\blpage{1027}
(\byear{2013})
\doiurl{10.1038/nmat3754}
\end{barticle}
\endbibitem

%%% 6
\bibitem[\protect\citeauthoryear{Hickox-Young et~al.}{2023}]{hockoxyoung2023}
\begin{barticle}
\bauthor{\bsnm{Hickox-Young}, \binits{D.}},
\bauthor{\bsnm{Puggioni}, \binits{D.}},
\bauthor{\bsnm{Rondinelli}, \binits{J.M.}}:
\batitle{Polar metals taxonomy for materials classification and discovery}.
\bjtitle{Physical Review Materials}
\bvolume{7},
\bfpage{010301}
(\byear{2023})
\doiurl{10.1103/PhysRevMaterials.7.010301}
\end{barticle}
\endbibitem

%%% 7
\bibitem[\protect\citeauthoryear{Ke et~al.}{2021}]{ke2021}
\begin{barticle}
\bauthor{\bsnm{Ke}, \binits{C.}},
\bauthor{\bsnm{Huang}, \binits{J.}},
\bauthor{\bsnm{Liu}, \binits{S.}}:
\batitle{Two-dimensional ferroelectric metal for electrocatalysis}.
\bjtitle{Mater. Horiz.}
\bvolume{8},
\bfpage{3387}--\blpage{3393}
(\byear{2021})
\doiurl{10.1039/D1MH01556G}
\end{barticle}
\endbibitem

%%% 8
\bibitem[\protect\citeauthoryear{Zhao et~al.}{2018}]{zhao2018}
\begin{barticle}
\bauthor{\bsnm{Zhao}, \binits{H.J.}},
\bauthor{\bsnm{Filippetti}, \binits{A.}},
\bauthor{\bsnm{Escorihuela-Sayalero}, \binits{C.}},
\bauthor{\bsnm{Delugas}, \binits{P.}},
\bauthor{\bsnm{Canadell}, \binits{E.}},
\bauthor{\bsnm{Bellaiche}, \binits{L.}},
\bauthor{\bsnm{Fiorentini}, \binits{V.}},
\bauthor{\bsnm{\'I\~niguez}, \binits{J.}}:
\batitle{Meta-screening and permanence of polar distortion in metallized ferroelectrics}.
\bjtitle{Physical Review B}
\bvolume{97},
\bfpage{054107}
(\byear{2018})
\doiurl{10.1103/PhysRevB.97.054107}
\end{barticle}
\endbibitem

%%% 9
\bibitem[\protect\citeauthoryear{Gu et~al.}{2017}]{Gu2017}
\begin{barticle}
\bauthor{\bsnm{Gu}, \binits{J.-x.}},
\bauthor{\bsnm{Jin}, \binits{K.-j.}},
\bauthor{\bsnm{Ma}, \binits{C.}},
\bauthor{\bsnm{Zhang}, \binits{Q.-h.}},
\bauthor{\bsnm{Gu}, \binits{L.}},
\bauthor{\bsnm{Ge}, \binits{C.}},
\bauthor{\bsnm{Wang}, \binits{J.-s.}},
\bauthor{\bsnm{Wang}, \binits{C.}},
\bauthor{\bsnm{Guo}, \binits{H.-z.}},
\bauthor{\bsnm{Yang}, \binits{G.-z.}}:
\batitle{Coexistence of polar distortion and metallicity in {P}b{T}i$_{1\text{\ensuremath{-}}x}${N}b$_{x}${O}$_{3}$}.
\bjtitle{Physical Review B}
\bvolume{96},
\bfpage{165206}
(\byear{2017})
\doiurl{10.1103/PhysRevB.96.165206}
\end{barticle}
\endbibitem

%%% 10
\bibitem[\protect\citeauthoryear{Cao et~al.}{2018}]{Cao2018}
\begin{botherref}
\oauthor{\bsnm{Cao}, \binits{Y.}},
\oauthor{\bsnm{Wang}, \binits{Z.}},
\oauthor{\bsnm{Park}, \binits{S.Y.}},
\oauthor{\bsnm{Yuan}, \binits{Y.}},
\oauthor{\bsnm{Liu}, \binits{X.}},
\oauthor{\bsnm{Nikitin}, \binits{S.M.}},
\oauthor{\bsnm{Akamatsu}, \binits{H.}},
\oauthor{\bsnm{Kareev}, \binits{M.}},
\oauthor{\bsnm{Middey}, \binits{S.}},
\oauthor{\bsnm{Meyers}, \binits{D.}},
\oauthor{\bsnm{Thompson}, \binits{P.}},
\oauthor{\bsnm{Ryan}, \binits{P.J.}},
\oauthor{\bsnm{Shafer}, \binits{P.}},
\oauthor{\bsnm{N’Diaye}, \binits{A.}},
\oauthor{\bsnm{Arenholz}, \binits{E.}},
\oauthor{\bsnm{Gopalan}, \binits{V.}},
\oauthor{\bsnm{Zhu}, \binits{Y.}},
\oauthor{\bsnm{Rabe}, \binits{K.M.}},
\oauthor{\bsnm{Chakhalian}, \binits{J.}}:
Artificial two-dimensional polar metal at room temperature.
Nature Communications
\textbf{9}(1)
(2018)
\doiurl{10.1038/s41467-018-03964-9}
\end{botherref}
\endbibitem

%%% 11
\bibitem[\protect\citeauthoryear{Lei et~al.}{2018}]{Lei2018}
\begin{barticle}
\bauthor{\bsnm{Lei}, \binits{S.}},
\bauthor{\bsnm{Gu}, \binits{M.}},
\bauthor{\bsnm{Puggioni}, \binits{D.}},
\bauthor{\bsnm{Stone}, \binits{G.}},
\bauthor{\bsnm{Peng}, \binits{J.}},
\bauthor{\bsnm{Ge}, \binits{J.}},
\bauthor{\bsnm{Wang}, \binits{Y.}},
\bauthor{\bsnm{Wang}, \binits{B.}},
\bauthor{\bsnm{Yuan}, \binits{Y.}},
\bauthor{\bsnm{Wang}, \binits{K.}},
\bauthor{\bsnm{Mao}, \binits{Z.}},
\bauthor{\bsnm{Rondinelli}, \binits{J.M.}},
\bauthor{\bsnm{Gopalan}, \binits{V.}}:
\batitle{Observation of quasi-two-dimensional polar domains and ferroelastic switching in a metal, {C}a$_3${R}u$_2${O}$_7$}.
\bjtitle{Nano Letters}
\bvolume{18}(\bissue{5}),
\bfpage{3088}--\blpage{3095}
(\byear{2018})
\doiurl{10.1021/acs.nanolett.8b00633}
\end{barticle}
\endbibitem

%%% 12
\bibitem[\protect\citeauthoryear{Fei et~al.}{2018}]{Fei2018}
\begin{barticle}
\bauthor{\bsnm{Fei}, \binits{Z.}},
\bauthor{\bsnm{Zhao}, \binits{W.}},
\bauthor{\bsnm{Palomaki}, \binits{T.A.}},
\bauthor{\bsnm{Sun}, \binits{B.}},
\bauthor{\bsnm{Miller}, \binits{M.K.}},
\bauthor{\bsnm{Zhao}, \binits{Z.}},
\bauthor{\bsnm{Yan}, \binits{J.}},
\bauthor{\bsnm{Xu}, \binits{X.}},
\bauthor{\bsnm{Cobden}, \binits{D.H.}}:
\batitle{Ferroelectric switching of a two-dimensional metal}.
\bjtitle{Nature}
\bvolume{560}(\bissue{7718}),
\bfpage{336}--\blpage{339}
(\byear{2018})
\doiurl{10.1038/s41586-018-0336-3}
\end{barticle}
\endbibitem

%%% 13
\bibitem[\protect\citeauthoryear{J\"{a}ger et~al.}{2024}]{Jager2024}
\begin{botherref}
\oauthor{\bsnm{J\"{a}ger}, \binits{F.}},
\oauthor{\bsnm{Spaldin}, \binits{N.A.}},
\oauthor{\bsnm{Bhowal}, \binits{S.}}:
Universal responses in nonmagnetic polar metals.
Physical Review Research
\textbf{6}(1)
(2024)
\doiurl{10.1103/physrevresearch.6.013251}
\end{botherref}
\endbibitem

%%% 14
\bibitem[\protect\citeauthoryear{Varjas et~al.}{2016}]{Varjas2016}
\begin{botherref}
\oauthor{\bsnm{Varjas}, \binits{D.}},
\oauthor{\bsnm{Grushin}, \binits{A.G.}},
\oauthor{\bsnm{Ilan}, \binits{R.}},
\oauthor{\bsnm{Moore}, \binits{J.E.}}:
Dynamical piezoelectric and magnetopiezoelectric effects in polar metals from berry phases and orbital moments.
Physical Review Letters
\textbf{117}(25)
(2016)
\doiurl{10.1103/physrevlett.117.257601}
\end{botherref}
\endbibitem

%%% 15
\bibitem[\protect\citeauthoryear{Padmanabhan et~al.}{2018}]{Padmanabhan2018}
\begin{botherref}
\oauthor{\bsnm{Padmanabhan}, \binits{H.}},
\oauthor{\bsnm{Park}, \binits{Y.}},
\oauthor{\bsnm{Puggioni}, \binits{D.}},
\oauthor{\bsnm{Yuan}, \binits{Y.}},
\oauthor{\bsnm{Cao}, \binits{Y.}},
\oauthor{\bsnm{Gasparov}, \binits{L.}},
\oauthor{\bsnm{Shi}, \binits{Y.}},
\oauthor{\bsnm{Chakhalian}, \binits{J.}},
\oauthor{\bsnm{Rondinelli}, \binits{J.M.}},
\oauthor{\bsnm{Gopalan}, \binits{V.}}:
Linear and nonlinear optical probe of the ferroelectric-like phase transition in a polar metal, {L}i{O}s{O}$_3$.
Applied Physics Letters
\textbf{113}(12)
(2018)
\doiurl{10.1063/1.5042769}
\end{botherref}
\endbibitem

%%% 16
\bibitem[\protect\citeauthoryear{Puggioni and Rondinelli}{2014}]{Puggioni2014}
\begin{botherref}
\oauthor{\bsnm{Puggioni}, \binits{D.}},
\oauthor{\bsnm{Rondinelli}, \binits{J.M.}}:
Designing a robustly metallic noncenstrosymmetric ruthenate oxide with large thermopower anisotropy.
Nature Communications
\textbf{5}(1)
(2014)
\doiurl{10.1038/ncomms4432}
\end{botherref}
\endbibitem

%%% 17
\bibitem[\protect\citeauthoryear{Hwang et~al.}{2025}]{Hwang2025}
\begin{barticle}
\bauthor{\bsnm{Hwang}, \binits{E.}},
\bauthor{\bsnm{Baek}, \binits{S.}},
\bauthor{\bsnm{Cho}, \binits{W.}},
\bauthor{\bsnm{Joo}, \binits{Y.}},
\bauthor{\bsnm{Jung}, \binits{J.}},
\bauthor{\bsnm{Watanabe}, \binits{K.}},
\bauthor{\bsnm{Taniguchi}, \binits{T.}},
\bauthor{\bsnm{Kim}, \binits{Y.-H.}},
\bauthor{\bsnm{Yang}, \binits{H.}}:
\batitle{Polar ohmic contact switching with a ferroelectric metal}.
\bjtitle{ACS Applied Materials \& Interfaces}
\bvolume{17}(\bissue{15}),
\bfpage{22984}--\blpage{22991}
(\byear{2025})
\doiurl{10.1021/acsami.5c02249}
\end{barticle}
\endbibitem

%%% 18
\bibitem[\protect\citeauthoryear{Catalan et~al.}{2012}]{Catalan2012}
\begin{barticle}
\bauthor{\bsnm{Catalan}, \binits{G.}},
\bauthor{\bsnm{Seidel}, \binits{J.}},
\bauthor{\bsnm{Ramesh}, \binits{R.}},
\bauthor{\bsnm{Scott}, \binits{J.F.}}:
\batitle{Domain wall nanoelectronics}.
\bjtitle{Reviews of Modern Physics}
\bvolume{84}(\bissue{1}),
\bfpage{119}--\blpage{156}
(\byear{2012})
\doiurl{10.1103/revmodphys.84.119}
\end{barticle}
\endbibitem

%%% 19
\bibitem[\protect\citeauthoryear{Stone et~al.}{2019}]{Stone2019}
\begin{botherref}
\oauthor{\bsnm{Stone}, \binits{G.}},
\oauthor{\bsnm{Puggioni}, \binits{D.}},
\oauthor{\bsnm{Lei}, \binits{S.}},
\oauthor{\bsnm{Gu}, \binits{M.}},
\oauthor{\bsnm{Wang}, \binits{K.}},
\oauthor{\bsnm{Wang}, \binits{Y.}},
\oauthor{\bsnm{Ge}, \binits{J.}},
\oauthor{\bsnm{Lu}, \binits{X.-Z.}},
\oauthor{\bsnm{Mao}, \binits{Z.}},
\oauthor{\bsnm{Rondinelli}, \binits{J.M.}},
\oauthor{\bsnm{Gopalan}, \binits{V.}}:
Atomic and electronic structure of domains walls in a polar metal.
Physical Review B
\textbf{99}(1)
(2019)
\doiurl{10.1103/physrevb.99.014105}
\end{botherref}
\endbibitem

%%% 20
\bibitem[\protect\citeauthoryear{Bednyakov et~al.}{2018}]{Bednyakov2018}
\begin{botherref}
\oauthor{\bsnm{Bednyakov}, \binits{P.S.}},
\oauthor{\bsnm{Sturman}, \binits{B.I.}},
\oauthor{\bsnm{Sluka}, \binits{T.}},
\oauthor{\bsnm{Tagantsev}, \binits{A.K.}},
\oauthor{\bsnm{Yudin}, \binits{P.V.}}:
Physics and applications of charged domain walls.
npj Computational Materials
\textbf{4}(1)
(2018)
\doiurl{10.1038/s41524-018-0121-8}
\end{botherref}
\endbibitem

%%% 21
\bibitem[\protect\citeauthoryear{Seidel et~al.}{2010}]{Seidel2010}
\begin{botherref}
\oauthor{\bsnm{Seidel}, \binits{J.}},
\oauthor{\bsnm{Maksymovych}, \binits{P.}},
\oauthor{\bsnm{Batra}, \binits{Y.}},
\oauthor{\bsnm{Katan}, \binits{A.}},
\oauthor{\bsnm{Yang}, \binits{S.-Y.}},
\oauthor{\bsnm{He}, \binits{Q.}},
\oauthor{\bsnm{Baddorf}, \binits{A.P.}},
\oauthor{\bsnm{Kalinin}, \binits{S.V.}},
\oauthor{\bsnm{Yang}, \binits{C.-H.}},
\oauthor{\bsnm{Yang}, \binits{J.-C.}},
\oauthor{\bsnm{Chu}, \binits{Y.-H.}},
\oauthor{\bsnm{Salje}, \binits{E.K.H.}},
\oauthor{\bsnm{Wormeester}, \binits{H.}},
\oauthor{\bsnm{Salmeron}, \binits{M.}},
\oauthor{\bsnm{Ramesh}, \binits{R.}}:
Domain wall conductivity in {L}a-doped {B}i{F}e{O}$_3$.
Physical Review Letters
\textbf{105}(19)
(2010)
\doiurl{10.1103/physrevlett.105.197603}
\end{botherref}
\endbibitem

%%% 22
\bibitem[\protect\citeauthoryear{Aird and Salje}{1999}]{Aird1998}
\begin{barticle}
\bauthor{\bsnm{Aird}, \binits{A.}},
\bauthor{\bsnm{Salje}, \binits{E.}}:
\batitle{Sheet superconductivity in twin walls: experimental evidence of {WO}$_{3-x}$}.
\bjtitle{Journal of Physics: Condensed Matter}
\bvolume{10},
\bfpage{377}
(\byear{1999})
\doiurl{10.1088/0953-8984/10/22/003}
\end{barticle}
\endbibitem

%%% 23
\bibitem[\protect\citeauthoryear{Yang et~al.}{2010}]{Yang2010}
\begin{barticle}
\bauthor{\bsnm{Yang}, \binits{S.Y.}},
\bauthor{\bsnm{Seidel}, \binits{J.}},
\bauthor{\bsnm{Byrnes}, \binits{S.J.}},
\bauthor{\bsnm{Shafer}, \binits{P.}},
\bauthor{\bsnm{Yang}, \binits{C.-H.}},
\bauthor{\bsnm{Rossell}, \binits{M.D.}},
\bauthor{\bsnm{Yu}, \binits{P.}},
\bauthor{\bsnm{Chu}, \binits{Y.-H.}},
\bauthor{\bsnm{Scott}, \binits{J.F.}},
\bauthor{\bsnm{Ager}, \binits{J.W.}},
\bauthor{\bsnm{Martin}, \binits{L.W.}},
\bauthor{\bsnm{Ramesh}, \binits{R.}}:
\batitle{Above-bandgap voltages from ferroelectric photovoltaic devices}.
\bjtitle{Nature Nanotechnology}
\bvolume{5}(\bissue{2}),
\bfpage{143}--\blpage{147}
(\byear{2010})
\doiurl{10.1038/nnano.2009.451}
\end{barticle}
\endbibitem

%%% 24
\bibitem[\protect\citeauthoryear{Seidel and Ramesh}{2020}]{Seidel2020}
\begin{bbook}
\bauthor{\bsnm{Seidel}, \binits{J.}},
\bauthor{\bsnm{Ramesh}, \binits{R.}}:
\bbtitle{Electronics Based on Domain Walls},
pp. \bfpage{340}--\blpage{350}.
\bpublisher{Oxford University Press},
\blocation{Oxford}
(\byear{2020}).
\doiurl{10.1093/oso/9780198862499.003.0015}
\end{bbook}
\endbibitem

%%% 25
\bibitem[\protect\citeauthoryear{Gu et~al.}{2023}]{Gu2023}
\begin{botherref}
\oauthor{\bsnm{Gu}, \binits{F.}},
\oauthor{\bsnm{Wang}, \binits{J.}},
\oauthor{\bsnm{Lang}, \binits{Z.-J.}},
\oauthor{\bsnm{Ku}, \binits{W.}}:
Quantum fluctuation of ferroelectric order in polar metals.
npj Quantum Materials
\textbf{8}(1)
(2023)
\doiurl{10.1038/s41535-023-00578-3}
\end{botherref}
\endbibitem

%%% 26
\bibitem[\protect\citeauthoryear{Ellner}{1990}]{Ellner1990}
\begin{barticle}
\bauthor{\bsnm{Ellner}, \binits{M.}}:
\batitle{The structure of the high-temperature phase {M}n{A}l(h) and the displacive transformation from {M}n{A}l(h) into {M}n$_5${A}l$_8$}.
\bjtitle{Metallurgical Transactions A}
\bvolume{21}(\bissue{6}),
\bfpage{1669}--\blpage{1672}
(\byear{1990})
\doiurl{10.1007/bf02672582}
\end{barticle}
\endbibitem

%%% 27
\bibitem[\protect\citeauthoryear{Liu et~al.}{1999}]{Liu1999}
\begin{barticle}
\bauthor{\bsnm{Liu}, \binits{X.J.}},
\bauthor{\bsnm{Ohnuma}, \binits{I.}},
\bauthor{\bsnm{Kainuma}, \binits{R.}},
\bauthor{\bsnm{Ishida}, \binits{K.}}:
\batitle{Thermodynamic assessment of the aluminum-manganese ({A}l-{M}n) binary phase diagram}.
\bjtitle{Journal of Phase Equilibria}
\bvolume{20}(\bissue{1}),
\bfpage{45}--\blpage{56}
(\byear{1999})
\doiurl{10.1361/105497199770335938}
\end{barticle}
\endbibitem

%%% 28
\bibitem[\protect\citeauthoryear{Bradley and Thewlis}{1926}]{bradley1926}
\begin{barticle}
\bauthor{\bsnm{Bradley}, \binits{A.J.}},
\bauthor{\bsnm{Thewlis}, \binits{J.}}:
\batitle{The structure of $\gamma$ -brass}.
\bjtitle{Proceedings of the Royal Society of London. Series A, Containing Papers of a Mathematical and Physical Character}
\bvolume{112}(\bissue{762}),
\bfpage{678}--\blpage{692}
(\byear{1926}).
Accessed 2025-08-08
\end{barticle}
\endbibitem

%%% 29
\bibitem[\protect\citeauthoryear{Zeng et~al.}{2018}]{Zeng2018}
\begin{barticle}
\bauthor{\bsnm{Zeng}, \binits{G.}},
\bauthor{\bsnm{Xian}, \binits{J.W.}},
\bauthor{\bsnm{Gourlay}, \binits{C.M.}}:
\batitle{Nucleation and growth crystallography of {A}l$_8${M}n$_5$ on ${B}_2$-{A}l({M}n, {F}e) in {AZ}91 magnesium alloys}.
\bjtitle{Acta Materialia}
\bvolume{153},
\bfpage{364}--\blpage{376}
(\byear{2018})
\doiurl{10.1016/j.actamat.2018.04.032}
\end{barticle}
\endbibitem

%%% 30
\bibitem[\protect\citeauthoryear{Swenson}{1997}]{Swenson97}
\begin{barticle}
\bauthor{\bsnm{Swenson}, \binits{D.}}:
\batitle{The ir3ge7 (d8f) structure: An electron phase related to y-brass}.
\bjtitle{Mat. Res. Soc. Symp. Proc.}
\bvolume{453},
\bfpage{367}--\blpage{372}
(\byear{1997})
\doiurl{10.1557/PROC-453-367}
\end{barticle}
\endbibitem

%%% 31
\bibitem[\protect\citeauthoryear{Ali et~al.}{2014}]{Ali2014}
\begin{barticle}
\bauthor{\bsnm{Ali}, \binits{M.N.}},
\bauthor{\bsnm{Xiong}, \binits{J.}},
\bauthor{\bsnm{Flynn}, \binits{S.}},
\bauthor{\bsnm{Tao}, \binits{J.}},
\bauthor{\bsnm{Gibson}, \binits{Q.D.}},
\bauthor{\bsnm{Schoop}, \binits{L.M.}},
\bauthor{\bsnm{Liang}, \binits{T.}},
\bauthor{\bsnm{Haldolaarachchige}, \binits{N.}},
\bauthor{\bsnm{Hirschberger}, \binits{M.}},
\bauthor{\bsnm{Ong}, \binits{N.P.}},
\bauthor{\bsnm{Cava}, \binits{R.J.}}:
\batitle{Large, non-saturating magnetoresistance in {W}{T}e$_2$}.
\bjtitle{Nature}
\bvolume{514}(\bissue{7521}),
\bfpage{205}--\blpage{208}
(\byear{2014})
\doiurl{10.1038/nature13763}
\end{barticle}
\endbibitem

%%% 32
\bibitem[\protect\citeauthoryear{Desai et~al.}{1984}]{Desai1984}
\begin{barticle}
\bauthor{\bsnm{Desai}, \binits{P.D.}},
\bauthor{\bsnm{James}, \binits{H.M.}},
\bauthor{\bsnm{Ho}, \binits{C.Y.}}:
\batitle{Electrical resistivity of aluminum and manganese}.
\bjtitle{Journal of Physical and Chemical Reference Data}
\bvolume{13}(\bissue{4}),
\bfpage{1131}--\blpage{1172}
(\byear{1984})
\doiurl{10.1063/1.555725}
\end{barticle}
\endbibitem

%%% 33
\bibitem[\protect\citeauthoryear{Zhang et~al.}{2001}]{ZHANG2001}
\begin{barticle}
\bauthor{\bsnm{Zhang}, \binits{W.J.}},
\bauthor{\bsnm{Reddy}, \binits{B.V.}},
\bauthor{\bsnm{Deevi}, \binits{S.C.}}:
\batitle{Physical properties of {T}i{A}l-base alloys}.
\bjtitle{Scripta Materialia}
\bvolume{45}(\bissue{6}),
\bfpage{645}--\blpage{651}
(\byear{2001})
\doiurl{10.1016/S1359-6462(01)01075-2}
\end{barticle}
\endbibitem

%%% 34
\bibitem[\protect\citeauthoryear{Shirai et~al.}{1995}]{SHIRAI1995}
\begin{barticle}
\bauthor{\bsnm{Shirai}, \binits{Y.}},
\bauthor{\bsnm{Masaki}, \binits{K.}},
\bauthor{\bsnm{Inoue}, \binits{T.}},
\bauthor{\bsnm{Nishitani}, \binits{S.R.}},
\bauthor{\bsnm{Yamaguchi}, \binits{M.}}:
\batitle{Electrical resistivity of {L}l$_2$ trialuminides containing 3d transition element}.
\bjtitle{Intermetallics}
\bvolume{3}(\bissue{5}),
\bfpage{381}--\blpage{388}
(\byear{1995})
\doiurl{10.1016/0966-9795(95)94256-E}
\end{barticle}
\endbibitem

%%% 35
\bibitem[\protect\citeauthoryear{Thimmaiah et~al.}{2017}]{Thimmaiah2017}
\begin{barticle}
\bauthor{\bsnm{Thimmaiah}, \binits{S.}},
\bauthor{\bsnm{Tener}, \binits{Z.}},
\bauthor{\bsnm{Lamichhane}, \binits{T.N.}},
\bauthor{\bsnm{Canfield}, \binits{P.C.}},
\bauthor{\bsnm{Miller}, \binits{G.J.}}:
\batitle{Crystal structure, homogeneity range and electronic structure of rhombohedral $\gamma$-{M}n$_5${A}l$_8$}.
\bjtitle{Zeitschrift f\"{u}r Kristallographie - Crystalline Materials}
\bvolume{232}(\bissue{7–9}),
\bfpage{601}--\blpage{610}
(\byear{2017})
\doiurl{10.1515/zkri-2017-0003}
\end{barticle}
\endbibitem

%%% 36
\bibitem[\protect\citeauthoryear{Hickox-Young}{2021}]{hickoxyoung2021}
\begin{botherref}
\oauthor{\bsnm{Hickox-Young}, \binits{D.}}:
Materials and models for coexisting metallic conductivity and broken inversion symmetry.
PhD thesis,
Northwestern University
(2021).
\url{https://www.proquest.com/dissertations-theses/materials-models-coexisting-metallic-conductivity/docview/2572570759/se-2}
\end{botherref}
\endbibitem

%%% 37
\bibitem[\protect\citeauthoryear{Bhattacharya}{2003}]{bhattacharya2003}
\begin{bbook}
\bauthor{\bsnm{Bhattacharya}, \binits{K.}}:
\bbtitle{Microstructure of Martensite: Why It Forms and How It Gives Rise to the Shape-Memory Effect. Vol 2}.
\bpublisher{Oxford University Press},
\blocation{Oxford}
(\byear{2003})
\end{bbook}
\endbibitem

%%% 38
\bibitem[\protect\citeauthoryear{Tsou et~al.}{2011}]{Tsou2011}
\begin{barticle}
\bauthor{\bsnm{Tsou}, \binits{N.T.}},
\bauthor{\bsnm{Pontis}, \binits{P.R.}},
\bauthor{\bsnm{Huber}, \binits{J.E.}}:
\batitle{Classification of laminate domain patterns in ferroelectrics}.
\bjtitle{Physical Review B}
\bvolume{83},
\bfpage{184120}
(\byear{2011})
\doiurl{10.1103/PhysRevB.83.184120}
\end{barticle}
\endbibitem

%%% 39
\bibitem[\protect\citeauthoryear{Liu and Tsou}{2016}]{liu2016}
\begin{barticle}
\bauthor{\bsnm{Liu}, \binits{T.-C.}},
\bauthor{\bsnm{Tsou}, \binits{N.-T.}}:
\batitle{The observation of microstructures in the trigonal shape memory alloys}.
\bjtitle{Coupled systems mechanics}
\bvolume{5},
\bfpage{329}--\blpage{340}
(\byear{2016})
\doiurl{10.12989/csm.2016.5.4.329}
\end{barticle}
\endbibitem

%%% 40
\bibitem[\protect\citeauthoryear{Ball and James}{1987}]{Ball1987}
\begin{barticle}
\bauthor{\bsnm{Ball}, \binits{J.M.}},
\bauthor{\bsnm{James}, \binits{R.D.}}:
\batitle{Fine phase mixtures as minimizers of energy}.
\bjtitle{Archive for Rational Mechanics and Analysis}
\bvolume{100}(\bissue{1}),
\bfpage{13}--\blpage{52}
(\byear{1987})
\doiurl{10.1007/bf00281246}
\end{barticle}
\endbibitem

%%% 41
\bibitem[\protect\citeauthoryear{Erhart}{2004}]{Erhart2004}
\begin{barticle}
\bauthor{\bsnm{Erhart}, \binits{J.}}:
\batitle{Domain wall orientations in ferroelastics and ferroelectrics}.
\bjtitle{Phase Transitions}
\bvolume{77}(\bissue{12}),
\bfpage{989}--\blpage{1074}
(\byear{2004})
\doiurl{10.1080/01411590410001710744}
\end{barticle}
\endbibitem

%%% 42
\bibitem[\protect\citeauthoryear{Tsou et~al.}{2015}]{Tsou2015}
\begin{barticle}
\bauthor{\bsnm{Tsou}, \binits{N.-T.}},
\bauthor{\bsnm{Chen}, \binits{C.-H.}},
\bauthor{\bsnm{Chen}, \binits{C.-S.}},
\bauthor{\bsnm{Wu}, \binits{S.-K.}}:
\batitle{Classification and analysis of trigonal martensite laminate twins in shape memory alloys}.
\bjtitle{Acta Materialia}
\bvolume{89},
\bfpage{193}--\blpage{204}
(\byear{2015})
\doiurl{10.1016/j.actamat.2015.02.006}
\end{barticle}
\endbibitem

%%% 43
\bibitem[\protect\citeauthoryear{Lee et~al.}{2015}]{Lee2015}
\begin{barticle}
\bauthor{\bsnm{Lee}, \binits{H.S.}},
\bauthor{\bsnm{Kim}, \binits{B.-S.}},
\bauthor{\bsnm{Cho}, \binits{C.-W.}},
\bauthor{\bsnm{Oh}, \binits{M.-W.}},
\bauthor{\bsnm{Min}, \binits{B.-K.}},
\bauthor{\bsnm{Park}, \binits{S.-D.}},
\bauthor{\bsnm{Lee}, \binits{H.-W.}}:
\batitle{Herringbone structure in gete-based thermoelectric materials}.
\bjtitle{Acta Materialia}
\bvolume{91},
\bfpage{83}--\blpage{90}
(\byear{2015})
\doiurl{10.1016/j.actamat.2015.03.015}
\end{barticle}
\endbibitem

%%% 44
\bibitem[\protect\citeauthoryear{Vermeulen et~al.}{2016}]{Vermeulen2016}
\begin{barticle}
\bauthor{\bsnm{Vermeulen}, \binits{P.A.}},
\bauthor{\bsnm{Kumar}, \binits{A.}},
\bauthor{\bsnm{Brink}, \binits{G.H.}},
\bauthor{\bsnm{Blake}, \binits{G.R.}},
\bauthor{\bsnm{Kooi}, \binits{B.J.}}:
\batitle{Unravelling the domain structures in {G}e{T}e and {L}a{A}l{O}$_3$}.
\bjtitle{Crystal Growth \& Design}
\bvolume{16}(\bissue{10}),
\bfpage{5915}--\blpage{5922}
(\byear{2016})
\doiurl{10.1021/acs.cgd.6b00960}
\end{barticle}
\endbibitem

%%% 45
\bibitem[\protect\citeauthoryear{Ricote et~al.}{1999}]{ricote1999}
\begin{barticle}
\bauthor{\bsnm{Ricote}, \binits{J.}},
\bauthor{\bsnm{Whatmore}, \binits{R.W.}},
\bauthor{\bsnm{Barber}, \binits{D.}}:
\batitle{Studies of the ferroelectric domain configuration and polarization of rhombohedral {PZT} ceramics}.
\bjtitle{Journal of Physics: Condensed Matter}
\bvolume{12},
\bfpage{323}
(\byear{1999})
\doiurl{10.1088/0953-8984/12/3/311}
\end{barticle}
\endbibitem

%%% 46
\bibitem[\protect\citeauthoryear{Sande et~al.}{1979}]{Sande1979}
\begin{barticle}
\bauthor{\bsnm{Sande}, \binits{M.V.}},
\bauthor{\bsnm{Landuyt}, \binits{J.V.}},
\bauthor{\bsnm{Amelinckx}, \binits{S.}}:
\batitle{Direct imaging of the structure and structural defects of rhombohedral $\gamma$-brasses}.
\bjtitle{Phys. Stat. Sol. a}
\bvolume{55},
\bfpage{41}--\blpage{49}
(\byear{1979})
\end{barticle}
\endbibitem

%%% 47
\bibitem[\protect\citeauthoryear{Cantwell et~al.}{2014}]{Cantwell2014}
\begin{barticle}
\bauthor{\bsnm{Cantwell}, \binits{P.R.}},
\bauthor{\bsnm{Tang}, \binits{M.}},
\bauthor{\bsnm{Dillon}, \binits{S.J.}},
\bauthor{\bsnm{Luo}, \binits{J.}},
\bauthor{\bsnm{Rohrer}, \binits{G.S.}},
\bauthor{\bsnm{Harmer}, \binits{M.P.}}:
\batitle{Grain boundary complexions}.
\bjtitle{Acta Materialia}
\bvolume{62},
\bfpage{1}--\blpage{48}
(\byear{2014})
\doiurl{10.1016/j.actamat.2013.07.037}
\end{barticle}
\endbibitem

%%% 48
\bibitem[\protect\citeauthoryear{Huang et~al.}{2011}]{Huang2011}
\begin{botherref}
\oauthor{\bsnm{Huang}, \binits{C.W.}},
\oauthor{\bsnm{Chen}, \binits{Z.H.}},
\oauthor{\bsnm{Wang}, \binits{J.}},
\oauthor{\bsnm{Sritharan}, \binits{T.}},
\oauthor{\bsnm{Chen}, \binits{L.}}:
Stability and crossover of 71$^{\circ}$ and 109$^{\circ}$ domains influenced by the film thickness and depolarization field in rhombohedral ferroelectric thin films.
Journal of Applied Physics
\textbf{110}(1)
(2011)
\doiurl{10.1063/1.3607977}
\end{botherref}
\endbibitem

%%% 49
\bibitem[\protect\citeauthoryear{Nordahl et~al.}{2024}]{Nordahl2024}
\begin{barticle}
\bauthor{\bsnm{Nordahl}, \binits{G.}},
\bauthor{\bsnm{Dagenborg}, \binits{S.}},
\bauthor{\bsnm{Sørhaug}, \binits{J.}},
\bauthor{\bsnm{Nord}, \binits{M.}}:
\batitle{Exploring deep learning models for 4{D}-{STEM}-{DPC} data processing}.
\bjtitle{Ultramicroscopy}
\bvolume{267},
\bfpage{114058}
(\byear{2024})
\doiurl{10.1016/j.ultramic.2024.114058}
\end{barticle}
\endbibitem

%%% 50
\bibitem[\protect\citeauthoryear{Conroy et~al.}{2020}]{Conroy2020}
\begin{barticle}
\bauthor{\bsnm{Conroy}, \binits{M.}},
\bauthor{\bsnm{Moore}, \binits{K.}},
\bauthor{\bsnm{O’Connell}, \binits{E.}},
\bauthor{\bsnm{Jones}, \binits{L.}},
\bauthor{\bsnm{Downing}, \binits{C.}},
\bauthor{\bsnm{Whamore}, \binits{R.}},
\bauthor{\bsnm{Gruverman}, \binits{A.}},
\bauthor{\bsnm{Gregg}, \binits{M.}},
\bauthor{\bsnm{Bangert}, \binits{U.}}:
\batitle{Probing the dynamics of topologically protected charged ferroelectric domain walls with the electron beam at the atomic scale}.
\bjtitle{Microscopy and Microanalysis}
\bvolume{26}(\bissue{S2}),
\bfpage{3030}--\blpage{3032}
(\byear{2020})
\doiurl{10.1017/s1431927620023594}
\end{barticle}
\endbibitem

%%% 51
\bibitem[\protect\citeauthoryear{Takamoto et~al.}{2024}]{Takamoto2024}
\begin{barticle}
\bauthor{\bsnm{Takamoto}, \binits{M.}},
\bauthor{\bsnm{Seki}, \binits{T.}},
\bauthor{\bsnm{Ikuhara}, \binits{Y.}},
\bauthor{\bsnm{Shibata}, \binits{N.}}:
\batitle{Diffraction contrast of ferroelectric domains in dpc stem images}.
\bjtitle{Microscopy}
\bvolume{73}(\bissue{5}),
\bfpage{422}--\blpage{429}
(\byear{2024})
\doiurl{10.1093/jmicro/dfae019}
\end{barticle}
\endbibitem

%%% 52
\bibitem[\protect\citeauthoryear{Lei}{2017}]{lei2017}
\begin{botherref}
\oauthor{\bsnm{Lei}, \binits{S.}}:
Coupled phenomena in domains and domain walls in complex polar oxides.
PhD thesis,
Penn State
(2017)
\end{botherref}
\endbibitem

%%% 53
\bibitem[\protect\citeauthoryear{Zhou et~al.}{2023}]{Zhou2023}
\begin{botherref}
\oauthor{\bsnm{Zhou}, \binits{X.}},
\oauthor{\bsnm{Ahmadian}, \binits{A.}},
\oauthor{\bsnm{Gault}, \binits{B.}},
\oauthor{\bsnm{Ophus}, \binits{C.}},
\oauthor{\bsnm{Liebscher}, \binits{C.H.}},
\oauthor{\bsnm{Dehm}, \binits{G.}},
\oauthor{\bsnm{Raabe}, \binits{D.}}:
Atomic motifs govern the decoration of grain boundaries by interstitial solutes.
Nature Communications
\textbf{14}(1)
(2023)
\doiurl{10.1038/s41467-023-39302-x}
\end{botherref}
\endbibitem

%%% 54
\bibitem[\protect\citeauthoryear{Seto and Ohtsuka}{2022}]{ReciPro2022}
\begin{barticle}
\bauthor{\bsnm{Seto}, \binits{Y.}},
\bauthor{\bsnm{Ohtsuka}, \binits{M.}}:
\batitle{Recipro: free and open-source multipurpose crystallographic software intetgrating a crystal model database and viewer, diffraction and microscopy simulators, and diffraction data analysis tools}.
\bjtitle{Journal of Applied Crystallography}
\bvolume{55},
\bfpage{397}--\blpage{410}
(\byear{2022})
\doiurl{10.1107/S1600576722000139}
\end{barticle}
\endbibitem

%%% 55
\bibitem[\protect\citeauthoryear{Shibata et~al.}{2015}]{Shibata2015}
\begin{botherref}
\oauthor{\bsnm{Shibata}, \binits{N.}},
\oauthor{\bsnm{Findlay}, \binits{S.D.}},
\oauthor{\bsnm{Sasaki}, \binits{H.}},
\oauthor{\bsnm{Matsumoto}, \binits{T.}},
\oauthor{\bsnm{Sawada}, \binits{H.}},
\oauthor{\bsnm{Kohno}, \binits{Y.}},
\oauthor{\bsnm{Otomo}, \binits{S.}},
\oauthor{\bsnm{Minato}, \binits{R.}},
\oauthor{\bsnm{Ikuhara}, \binits{Y.}}:
Imaging of built-in electric field at a p-n junction by scanning transmission electron microscopy.
Scientific Reports
\textbf{5}(1)
(2015)
\doiurl{10.1038/srep10040}
\end{botherref}
\endbibitem

%%% 56
\bibitem[\protect\citeauthoryear{Strauch et~al.}{2023}]{Strauch2023}
\begin{barticle}
\bauthor{\bsnm{Strauch}, \binits{A.}},
\bauthor{\bsnm{M\"{a}rz}, \binits{B.}},
\bauthor{\bsnm{Denneulin}, \binits{T.}},
\bauthor{\bsnm{Cattaneo}, \binits{M.}},
\bauthor{\bsnm{Rosenauer}, \binits{A.}},
\bauthor{\bsnm{M\"{u}ller-Caspary}, \binits{K.}}:
\batitle{Systematic errors of electric field measurements in ferroelectrics by unit cell averaged momentum transfers in stem}.
\bjtitle{Microscopy and Microanalysis}
\bvolume{29}(\bissue{2}),
\bfpage{499}--\blpage{511}
(\byear{2023})
\doiurl{10.1093/micmic/ozad016}
\end{barticle}
\endbibitem

%%% 57
\bibitem[\protect\citeauthoryear{Hunnestad et~al.}{2020}]{Hunnestad2020}
\begin{botherref}
\oauthor{\bsnm{Hunnestad}, \binits{K.A.}},
\oauthor{\bsnm{Roede}, \binits{E.D.}},
\oauthor{\bsnm{Helvoort}, \binits{A.T.J.}},
\oauthor{\bsnm{Meier}, \binits{D.}}:
Characterization of ferroelectric domain walls by scanning electron microscopy.
Journal of Applied Physics
\textbf{128}(19)
(2020)
\doiurl{10.1063/5.0029284}
\end{botherref}
\endbibitem

%%% 58
\bibitem[\protect\citeauthoryear{Bellitto}{2012}]{Bellitto_2012}
\begin{bbook}
\bauthor{\bsnm{Bellitto}, \binits{V.}}:
\bbtitle{Atomic Force Microscopy}.
\bpublisher{IntechOpen},
\blocation{Rijeka}
(\byear{2012}).
\doiurl{10.5772/2673} .
\burl{https://doi.org/10.5772/2673}
\end{bbook}
\endbibitem

%%% 59
\bibitem[\protect\citeauthoryear{Lei et~al.}{2004}]{Lei2004}
\begin{barticle}
\bauthor{\bsnm{Lei}, \binits{C.H.}},
\bauthor{\bsnm{Das}, \binits{A.}},
\bauthor{\bsnm{Elliott}, \binits{M.}},
\bauthor{\bsnm{Macdonald}, \binits{J.E.}}:
\batitle{Quantitative electrostatic force microscopy-phase measurements}.
\bjtitle{Nanotechnology}
\bvolume{15}(\bissue{5}),
\bfpage{627}--\blpage{634}
(\byear{2004})
\doiurl{10.1088/0957-4484/15/5/038}
\end{barticle}
\endbibitem

%%% 60
\bibitem[\protect\citeauthoryear{Djokic}{1996}]{Djokic1996}
\begin{barticle}
\bauthor{\bsnm{Djokic}, \binits{S.S.}}:
\batitle{Cementation of copper on aluminum in alkaline solutions}.
\bjtitle{J. Electrochem. Soc.}
\bvolume{143},
\bfpage{1300}--\blpage{1305}
(\byear{1996})
\doiurl{10.1149/1.1836634}
\end{barticle}
\endbibitem

%%% 61
\bibitem[\protect\citeauthoryear{Searson et~al.}{1989}]{searson1989}
\begin{barticle}
\bauthor{\bsnm{Searson}, \binits{P.}},
\bauthor{\bsnm{Nagarkar}, \binits{P.}},
\bauthor{\bsnm{Laianision}, \binits{R.}}:
\batitle{The effect of density of states, work function and exchange integral of polycrystalline and single crystal surfaces on the hydrogen evolution reaction}.
\bjtitle{International Journal of Hydrogen Energy}
\bvolume{14}(\bissue{2}),
\bfpage{131}--\blpage{136}
(\byear{1989})
\doiurl{10.1016/0360-3199(89)90002-5}
\end{barticle}
\endbibitem

%%% 62
\bibitem[\protect\citeauthoryear{ping Liu et~al.}{2000}]{LIU2000}
\begin{barticle}
\bauthor{\bsnm{Liu}, \binits{H.-p.}},
\bauthor{\bsnm{Colarieti-Tosti}, \binits{M.}},
\bauthor{\bsnm{Broddefalk}, \binits{A.}},
\bauthor{\bsnm{Andersson}, \binits{Y.}},
\bauthor{\bsnm{Lidström}, \binits{E.}},
\bauthor{\bsnm{Eriksson}, \binits{O.}}:
\batitle{On the structural polymorphism of {C}e{P}t$_2${S}n$_2$: experiment and theory}.
\bjtitle{Journal of Alloys and Compounds}
\bvolume{306}(\bissue{1}),
\bfpage{30}--\blpage{39}
(\byear{2000})
\doiurl{10.1016/S0925-8388(00)00768-4}
\end{barticle}
\endbibitem

%%% 63
\bibitem[\protect\citeauthoryear{Gschneidner and Calderwood}{1988}]{Gschneidner1988}
\begin{barticle}
\bauthor{\bsnm{Gschneidner}, \binits{K.A.}},
\bauthor{\bsnm{Calderwood}, \binits{F.W.}}:
\batitle{The {A}l-{C}e (aluminum-cerium) system}.
\bjtitle{Bulletin of Alloy Phase Diagrams}
\bvolume{9}(\bissue{6}),
\bfpage{669}--\blpage{672}
(\byear{1988})
\doiurl{10.1007/bf02883162}
\end{barticle}
\endbibitem

%%% 64
\bibitem[\protect\citeauthoryear{Duggin}{1966}]{DUGGIN1966}
\begin{barticle}
\bauthor{\bsnm{Duggin}, \binits{M.J.}}:
\batitle{Further studies of martensitic transformations in gold-copper-zinc and copper-aluminium-nickel alloys}.
\bjtitle{Acta Metallurgica}
\bvolume{14}(\bissue{2}),
\bfpage{123}--\blpage{129}
(\byear{1966})
\doiurl{10.1016/0001-6160(66)90293-8}
\end{barticle}
\endbibitem

%%% 65
\bibitem[\protect\citeauthoryear{Kerber et~al.}{1998}]{kerber1998}
\begin{botherref}
\oauthor{\bsnm{Kerber}, \binits{H.}},
\oauthor{\bsnm{Deiseroth}, \binits{H.-J.}},
\oauthor{\bsnm{Walther}, \binits{R.}}:
Crystal structure of $\alpha$-tripotassium bismuthide, $\alpha$-{K}$_3${B}i, a revision.
Zeitschrift für Kristallographie - New Crystal Structures
\textbf{213}
(1998)
\doiurl{10.1524/ncrs.1998.213.14.501}
\end{botherref}
\endbibitem

%%% 66
\bibitem[\protect\citeauthoryear{Steurer}{2007}]{Steurer2007}
\begin{barticle}
\bauthor{\bsnm{Steurer}, \binits{W.}}:
\batitle{The samson phase, $\beta$-{M}g$_2${A}l$_3$, revisited}.
\bjtitle{Zeitschrift f\"{u}r Kristallographie}
\bvolume{222}(\bissue{6}),
\bfpage{259}--\blpage{288}
(\byear{2007})
\doiurl{10.1524/zkri.2007.222.6.259}
\end{barticle}
\endbibitem

%%% 67
\bibitem[\protect\citeauthoryear{Nayeb-Hashemi and Clark}{1985}]{NayebHashemi1985}
\begin{barticle}
\bauthor{\bsnm{Nayeb-Hashemi}, \binits{A.A.}},
\bauthor{\bsnm{Clark}, \binits{J.B.}}:
\batitle{The {I}n-{M}g (indium-magnesium) system}.
\bjtitle{Bulletin of Alloy Phase Diagrams}
\bvolume{6}(\bissue{2}),
\bfpage{149}--\blpage{160}
(\byear{1985})
\doiurl{10.1007/bf02869233}
\end{barticle}
\endbibitem

%%% 68
\bibitem[\protect\citeauthoryear{Nover and Schubert}{1980}]{NOVER1980}
\begin{barticle}
\bauthor{\bsnm{Nover}, \binits{G.}},
\bauthor{\bsnm{Schubert}, \binits{K.}}:
\batitle{The crystal structure of {Ni}{Z}n$_3$.r}.
\bjtitle{Journal of the Less Common Metals}
\bvolume{75}(\bissue{1}),
\bfpage{51}--\blpage{63}
(\byear{1980})
\doiurl{10.1016/0022-5088(80)90368-9}
\end{barticle}
\endbibitem

%%% 69
\bibitem[\protect\citeauthoryear{Gulay and Harbrecht}{2003}]{Gulay2003}
\begin{barticle}
\bauthor{\bsnm{Gulay}, \binits{L.D.}},
\bauthor{\bsnm{Harbrecht}, \binits{B.}}:
\batitle{The crystal structure of $\zeta_2$‐{A}l$_3${C}u$_4$‐$\delta$}.
\bjtitle{Zeitschrift f\"{u}r anorganische und allgemeine Chemie}
\bvolume{629}(\bissue{3}),
\bfpage{463}--\blpage{466}
(\byear{2003})
\doiurl{10.1002/zaac.200390076}
\end{barticle}
\endbibitem

%%% 70
\bibitem[\protect\citeauthoryear{Goedecke~T}{1996}]{goedecke1996}
\begin{botherref}
\oauthor{\bsnm{Goedecke~T}, \binits{E.M.}}:
Phase equilibria in the aluminum-rich portion of the binary system {C}o-{A}l and in the cobalt/aluminium-rich portion of the ternary system {C}o-{N}i-{A}l.
Z Metallkd
\textbf{11}(86)
(1996)
\end{botherref}
\endbibitem

%%% 71
\bibitem[\protect\citeauthoryear{Iwasaki}{1965}]{iwasaki1965}
\begin{barticle}
\bauthor{\bsnm{Iwasaki}, \binits{H.}}:
\batitle{The crystal structure and the phase transition of a metastable phase in the {A}u-37.8\% {Z}n alloy}.
\bjtitle{Journal of the Physical Society of Japan}
\bvolume{20}(\bissue{12}),
\bfpage{2129}--\blpage{2140}
(\byear{1965})
\doiurl{10.1143/JPSJ.20.2129}
\end{barticle}
\endbibitem

%%% 72
\bibitem[\protect\citeauthoryear{Zaremba et~al.}{2004}]{Zaremba2004}
\begin{barticle}
\bauthor{\bsnm{Zaremba}, \binits{V.I.}},
\bauthor{\bsnm{Kaczorowski}, \binits{D.}},
\bauthor{\bsnm{Rodewald}, \binits{U.C.}},
\bauthor{\bsnm{Hoffmann}, \binits{R.-D.}},
\bauthor{\bsnm{P\"{o}ttgen}, \binits{R.}}:
\batitle{{L}a{P}d{I}n$_2$ with {M}g{C}u{A}l$_2$ and {REP}d{I}n$_2$ ({RE} = {Y}, {P}r, {N}d, {S}m, {G}d-{T}m, {L}u) with {H}f{N}i{G}a$_2$-type structure: Synthesis, structure, and physical properties}.
\bjtitle{Chemistry of Materials}
\bvolume{16}(\bissue{3}),
\bfpage{466}--\blpage{476}
(\byear{2004})
\doiurl{10.1021/cm031139m}
\end{barticle}
\endbibitem

%%% 73
\bibitem[\protect\citeauthoryear{Lawson}{1978}]{lawson1978}
\begin{barticle}
\bauthor{\bsnm{Lawson}, \binits{A.C.}}:
\batitle{Low-temperature crystal structures and superconductivity of {H}f$_{1-x}${Z}r$_x$ alloys}.
\bjtitle{Physical Review B}
\bvolume{17},
\bfpage{1136}--\blpage{1138}
(\byear{1978})
\doiurl{10.1103/PhysRevB.17.1136}
\end{barticle}
\endbibitem

%%% 74
\bibitem[\protect\citeauthoryear{Bauer et~al.}{2004}]{Bauer2004}
\begin{botherref}
\oauthor{\bsnm{Bauer}, \binits{E.}},
\oauthor{\bsnm{Hilscher}, \binits{G.}},
\oauthor{\bsnm{Michor}, \binits{H.}},
\oauthor{\bsnm{Paul}, \binits{C.}},
\oauthor{\bsnm{Scheidt}, \binits{E.W.}},
\oauthor{\bsnm{Gribanov}, \binits{A.}},
\oauthor{\bsnm{Seropegin}, \binits{Y.}},
\oauthor{\bsnm{Noël}, \binits{H.}},
\oauthor{\bsnm{Sigrist}, \binits{M.}},
\oauthor{\bsnm{Rogl}, \binits{P.}}:
Heavy fermion superconductivity and magnetic order in noncentrosymmetric {C}e{P}t$_3${S}i.
Physical Review Letters
\textbf{92}(2)
(2004)
\doiurl{10.1103/physrevlett.92.027003}
\end{botherref}
\endbibitem

%%% 75
\bibitem[\protect\citeauthoryear{Yip}{2014}]{Yip2014}
\begin{barticle}
\bauthor{\bsnm{Yip}, \binits{S.}}:
\batitle{Noncentrosymmetric superconductors}.
\bjtitle{Annual Review of Condensed Matter Physics}
\bvolume{5}(\bissue{1}),
\bfpage{15}--\blpage{33}
(\byear{2014})
\doiurl{10.1146/annurev-conmatphys-031113-133912}
\end{barticle}
\endbibitem

%%% 76
\bibitem[\protect\citeauthoryear{Bulaevskii et~al.}{1976}]{bulaevskii1976}
\begin{botherref}
\oauthor{\bsnm{Bulaevskii}, \binits{L.N.}},
\oauthor{\bsnm{Guseinov}, \binits{A.A.}},
\oauthor{\bsnm{Rusinov}, \binits{A.I.}}:
Superconductivity in crystals without symmetry centers.
Sov. Phys. - JETP (Engl. Transl.); (United States)
\textbf{71:6}
(1976)
\end{botherref}
\endbibitem

%%% 77
\bibitem[\protect\citeauthoryear{Shi et~al.}{2024}]{Shi2024}
\begin{botherref}
\oauthor{\bsnm{Shi}, \binits{J.}},
\oauthor{\bsnm{You}, \binits{W.}},
\oauthor{\bsnm{Li}, \binits{X.}},
\oauthor{\bsnm{Gao}, \binits{F.Y.}},
\oauthor{\bsnm{Peng}, \binits{X.}},
\oauthor{\bsnm{Zhang}, \binits{S.}},
\oauthor{\bsnm{Li}, \binits{J.}},
\oauthor{\bsnm{Zhang}, \binits{Y.}},
\oauthor{\bsnm{Fu}, \binits{L.}},
\oauthor{\bsnm{Taylor}, \binits{P.J.}},
\oauthor{\bsnm{Nelson}, \binits{K.A.}},
\oauthor{\bsnm{Baldini}, \binits{E.}}:
Revealing a distortive polar order buried in the {F}ermi sea.
Science Advances
\textbf{10}(28)
(2024)
\doiurl{10.1126/sciadv.adn0929}
\end{botherref}
\endbibitem

%%% 78
\bibitem[\protect\citeauthoryear{Princep et~al.}{2020}]{Princep2020}
\begin{botherref}
\oauthor{\bsnm{Princep}, \binits{A.J.}},
\oauthor{\bsnm{Feng}, \binits{H.L.}},
\oauthor{\bsnm{Guo}, \binits{Y.F.}},
\oauthor{\bsnm{Lang}, \binits{F.}},
\oauthor{\bsnm{Weng}, \binits{H.M.}},
\oauthor{\bsnm{Manuel}, \binits{P.}},
\oauthor{\bsnm{Khalyavin}, \binits{D.}},
\oauthor{\bsnm{Senyshyn}, \binits{A.}},
\oauthor{\bsnm{Rahn}, \binits{M.C.}},
\oauthor{\bsnm{Yuan}, \binits{Y.H.}},
\oauthor{\bsnm{Matsushita}, \binits{Y.}},
\oauthor{\bsnm{Blundell}, \binits{S.J.}},
\oauthor{\bsnm{Yamaura}, \binits{K.}},
\oauthor{\bsnm{Boothroyd}, \binits{A.T.}}:
Magnetically driven loss of centrosymmetry in metallic {P}b$_2${C}o{O}s{O}$_6$.
Physical Review B
\textbf{102}(10)
(2020)
\doiurl{10.1103/physrevb.102.104410}
\end{botherref}
\endbibitem

%%% 79
\bibitem[\protect\citeauthoryear{Benedek and Birol}{2016}]{Benedek2016}
\begin{barticle}
\bauthor{\bsnm{Benedek}, \binits{N.A.}},
\bauthor{\bsnm{Birol}, \binits{T.}}:
\batitle{‘{F}erroelectric’ metals reexamined: fundamental mechanisms and design considerations for new materials}.
\bjtitle{Journal of Materials Chemistry C}
\bvolume{4}(\bissue{18}),
\bfpage{4000}--\blpage{4015}
(\byear{2016})
\doiurl{10.1039/c5tc03856a}
\end{barticle}
\endbibitem

%%% 80
\bibitem[\protect\citeauthoryear{Íñiguez}{2020}]{iguez2020}
\begin{bbook}
\bauthor{\bsnm{Íñiguez}, \binits{J.}}:
\bbtitle{First-Principles Studies of Structural Domain Walls},
pp. \bfpage{36}--\blpage{75}.
\bpublisher{Oxford University Press},
\blocation{Oxford}
(\byear{2020}).
\doiurl{10.1093/oso/9780198862499.003.0003}
\end{bbook}
\endbibitem

%%% 81
\bibitem[\protect\citeauthoryear{Hanson et~al.}{2006}]{Hanson2006}
\begin{barticle}
\bauthor{\bsnm{Hanson}, \binits{J.N.}},
\bauthor{\bsnm{Rodriguez}, \binits{B.J.}},
\bauthor{\bsnm{Nemanich}, \binits{R.J.}},
\bauthor{\bsnm{Gruverman}, \binits{A.}}:
\batitle{Fabrication of metallic nanowires on a ferroelectric template via photochemical reaction}.
\bjtitle{Nanotechnology}
\bvolume{17}(\bissue{19}),
\bfpage{4946}--\blpage{4949}
(\byear{2006})
\doiurl{10.1088/0957-4484/17/19/028}
\end{barticle}
\endbibitem

%%% 82
\bibitem[\protect\citeauthoryear{Burbure et~al.}{2010}]{Burbure2010}
\begin{barticle}
\bauthor{\bsnm{Burbure}, \binits{N.V.}},
\bauthor{\bsnm{Salvador}, \binits{P.A.}},
\bauthor{\bsnm{Rohrer}, \binits{G.S.}}:
\batitle{Photochemical reactivity of titania films on {B}a{T}i{O}$_3$ substrates: Origin of spatial selectivity}.
\bjtitle{Chemistry of Materials}
\bvolume{22}(\bissue{21}),
\bfpage{5823}--\blpage{5830}
(\byear{2010})
\doiurl{10.1021/cm1018025}
\end{barticle}
\endbibitem

%%% 83
\bibitem[\protect\citeauthoryear{Lin et~al.}{2023}]{Lin2023}
\begin{botherref}
\oauthor{\bsnm{Lin}, \binits{L.}},
\oauthor{\bsnm{Jacobs}, \binits{R.}},
\oauthor{\bsnm{Ma}, \binits{T.}},
\oauthor{\bsnm{Chen}, \binits{D.}},
\oauthor{\bsnm{Booske}, \binits{J.}},
\oauthor{\bsnm{Morgan}, \binits{D.}}:
Work function: Fundamentals, measurement, calculation, engineering, and applications.
Physical Review Applied
\textbf{19}(3)
(2023)
\doiurl{10.1103/physrevapplied.19.037001}
\end{botherref}
\endbibitem

%%% 84
\bibitem[\protect\citeauthoryear{Chen et~al.}{2024}]{Chen2024}
\begin{botherref}
\oauthor{\bsnm{Chen}, \binits{Z.}},
\oauthor{\bsnm{Ma}, \binits{T.}},
\oauthor{\bsnm{Wei}, \binits{W.}},
\oauthor{\bsnm{Wong}, \binits{W.Y.}},
\oauthor{\bsnm{Zhao}, \binits{C.}},
\oauthor{\bsnm{Ni}, \binits{B.J.}}:
Work function-guided electrocatalyst design.
Advanced Materials
\textbf{36}
(2024)
\end{botherref}
\endbibitem

%%% 85
\bibitem[\protect\citeauthoryear{Bard et~al.}{1985}]{StandardPotentials}
\begin{bbook}
\bauthor{\bsnm{Bard}, \binits{A.J.}},
\bauthor{\bsnm{Parsons}, \binits{R.}},
\bauthor{\bsnm{Jordan}, \binits{J.E.}}:
\bbtitle{Standard Potentials in Aqueous Solution}.
\bpublisher{CRC Press}, \blocation{???}
(\byear{1985})
\end{bbook}
\endbibitem

%%% 86
\bibitem[\protect\citeauthoryear{Khisamov et~al.}{2013}]{Khisamov2013}
\begin{barticle}
\bauthor{\bsnm{Khisamov}, \binits{R.K.}},
\bauthor{\bsnm{Safarov}, \binits{I.M.}},
\bauthor{\bsnm{Mulyukov}, \binits{R.R.}},
\bauthor{\bsnm{Yumaguzin}, \binits{Y.M.}}:
\batitle{Effect of grain boundaries on the electron work function of nanocrystalline nickel}.
\bjtitle{Physics of the Solid State}
\bvolume{55}(\bissue{1}),
\bfpage{1}--\blpage{4}
(\byear{2013})
\doiurl{10.1134/s1063783413010186}
\end{barticle}
\endbibitem

%%% 87
\bibitem[\protect\citeauthoryear{Orlova et~al.}{2018}]{Orlova2018}
\begin{barticle}
\bauthor{\bsnm{Orlova}, \binits{T.S.}},
\bauthor{\bsnm{Ankudinov}, \binits{A.V.}},
\bauthor{\bsnm{Mavlyutov}, \binits{A.M.}},
\bauthor{\bsnm{Resnina}, \binits{N.N.}}:
\batitle{Effect of grain boundaries on the electron work function of ultrafine grained aluminum}.
\bjtitle{Reviews on Advanced Materials Science}
\bvolume{57}(\bissue{1}),
\bfpage{110}--\blpage{115}
(\byear{2018})
\doiurl{10.1515/rams-2018-0053}
\end{barticle}
\endbibitem

%%% 88
\bibitem[\protect\citeauthoryear{Hutchison~TS}{1968}]{hutchinson1968}
\begin{bbook}
\bauthor{\bsnm{Hutchison~TS}, \binits{B.D.}}:
\bbtitle{The Physics of Engineering Solids, Second Edition}.
\bpublisher{John Wiley, and Sons, Inc},
\blocation{New Jersey}
(\byear{1968})
\end{bbook}
\endbibitem

%%% 89
\bibitem[\protect\citeauthoryear{Ni et~al.}{2024}]{ni2024}
\begin{barticle}
\bauthor{\bsnm{Ni}, \binits{J.}},
\bauthor{\bsnm{Cao}, \binits{L.}},
\bauthor{\bsnm{Zhong}, \binits{B.}},
\bauthor{\bsnm{Li}, \binits{Q.}},
\bauthor{\bsnm{Guo}, \binits{C.}},
\bauthor{\bsnm{Song}, \binits{J.}},
\bauthor{\bsnm{Liu}, \binits{Y.}},
\bauthor{\bsnm{Lu}, \binits{M.}},
\bauthor{\bsnm{Fan}, \binits{T.}}:
\batitle{Characterizing local electronic states of twin boundaries in copper}.
\bjtitle{Nano Letters}
\bvolume{24}(\bissue{18}),
\bfpage{5474}--\blpage{5480}
(\byear{2024})
\doiurl{10.1021/acs.nanolett.4c00550} .
\bcomment{PMID: 38652833}
\end{barticle}
\endbibitem

%%% 90
\bibitem[\protect\citeauthoryear{Paxton et~al.}{1997}]{Pettifor1997}
\begin{barticle}
\bauthor{\bsnm{Paxton}, \binits{A.T.}},
\bauthor{\bsnm{Methfessel}, \binits{M.}},
\bauthor{\bsnm{Pettifor}, \binits{D.G.}}:
\batitle{A bandstructure view of the hume-rothery electron phases}.
\bjtitle{Proc. R. Soc. Lond. A}
\bvolume{453},
\bfpage{1493}--\blpage{1514}
(\byear{1997})
\end{barticle}
\endbibitem

%%% 91
\bibitem[\protect\citeauthoryear{Kim et~al.}{2016}]{Kim2016}
\begin{barticle}
\bauthor{\bsnm{Kim}, \binits{T.H.}},
\bauthor{\bsnm{Puggioni}, \binits{D.}},
\bauthor{\bsnm{Yuan}, \binits{Y.}},
\bauthor{\bsnm{Xie}, \binits{L.}},
\bauthor{\bsnm{Zhou}, \binits{H.}},
\bauthor{\bsnm{Campbell}, \binits{N.}},
\bauthor{\bsnm{Ryan}, \binits{P.J.}},
\bauthor{\bsnm{Choi}, \binits{Y.}},
\bauthor{\bsnm{Kim}, \binits{J.-W.}},
\bauthor{\bsnm{Patzner}, \binits{J.R.}},
\bauthor{\bsnm{Ryu}, \binits{S.}},
\bauthor{\bsnm{Podkaminer}, \binits{J.P.}},
\bauthor{\bsnm{Irwin}, \binits{J.}},
\bauthor{\bsnm{Ma}, \binits{Y.}},
\bauthor{\bsnm{Fennie}, \binits{C.J.}},
\bauthor{\bsnm{Rzchowski}, \binits{M.S.}},
\bauthor{\bsnm{Pan}, \binits{X.Q.}},
\bauthor{\bsnm{Gopalan}, \binits{V.}},
\bauthor{\bsnm{Rondinelli}, \binits{J.M.}},
\bauthor{\bsnm{Eom}, \binits{C.B.}}:
\batitle{Polar metals by geometric design}.
\bjtitle{Nature}
\bvolume{533}(\bissue{7601}),
\bfpage{68}--\blpage{72}
(\byear{2016})
\doiurl{10.1038/nature17628}
\end{barticle}
\endbibitem

%%% 92
\bibitem[\protect\citeauthoryear{Laurita et~al.}{2019}]{Laurita2019}
\begin{botherref}
\oauthor{\bsnm{Laurita}, \binits{N.J.}},
\oauthor{\bsnm{Ron}, \binits{A.}},
\oauthor{\bsnm{Shan}, \binits{J.-Y.}},
\oauthor{\bsnm{Puggioni}, \binits{D.}},
\oauthor{\bsnm{Koocher}, \binits{N.Z.}},
\oauthor{\bsnm{Yamaura}, \binits{K.}},
\oauthor{\bsnm{Shi}, \binits{Y.}},
\oauthor{\bsnm{Rondinelli}, \binits{J.M.}},
\oauthor{\bsnm{Hsieh}, \binits{D.}}:
Evidence for the weakly coupled electron mechanism in an anderson-blount polar metal.
Nature Communications
\textbf{10}(1)
(2019)
\doiurl{10.1038/s41467-019-11172-2}
\end{botherref}
\endbibitem

%%% 93
\bibitem[\protect\citeauthoryear{Lo~Vecchio et~al.}{2016}]{LoVecchio2016}
\begin{botherref}
\oauthor{\bsnm{Lo~Vecchio}, \binits{I.}},
\oauthor{\bsnm{Giovannetti}, \binits{G.}},
\oauthor{\bsnm{Autore}, \binits{M.}},
\oauthor{\bsnm{Di~Pietro}, \binits{P.}},
\oauthor{\bsnm{Perucchi}, \binits{A.}},
\oauthor{\bsnm{He}, \binits{J.}},
\oauthor{\bsnm{Yamaura}, \binits{K.}},
\oauthor{\bsnm{Capone}, \binits{M.}},
\oauthor{\bsnm{Lupi}, \binits{S.}}:
Electronic correlations in the ferroelectric metallic state of {L}i{O}s{O}$_3$.
Physical Review B
\textbf{93}(16)
(2016)
\doiurl{10.1103/physrevb.93.161113}
\end{botherref}
\endbibitem

%%% 94
\bibitem[\protect\citeauthoryear{Hlinka et~al.}{2016}]{Hlinka2016}
\begin{botherref}
\oauthor{\bsnm{Hlinka}, \binits{J.}},
\oauthor{\bsnm{Privratska}, \binits{J.}},
\oauthor{\bsnm{Ondrejkovic}, \binits{P.}},
\oauthor{\bsnm{Janovec}, \binits{V.}}:
Symmetry guide to ferroaxial transitions.
Physical Review Letters
\textbf{116}(17)
(2016)
\doiurl{10.1103/physrevlett.116.177602}
\end{botherref}
\endbibitem

%%% 95
\bibitem[\protect\citeauthoryear{Zabalo and Stengel}{2021}]{Zabalo2021}
\begin{botherref}
\oauthor{\bsnm{Zabalo}, \binits{A.}},
\oauthor{\bsnm{Stengel}, \binits{M.}}:
Switching a polar metal via strain gradients.
Physical Review Letters
\textbf{126}(12)
(2021)
\doiurl{10.1103/physrevlett.126.127601}
\end{botherref}
\endbibitem

%%% 96
\bibitem[\protect\citeauthoryear{Peng et~al.}{2024}]{peng2024}
\begin{barticle}
\bauthor{\bsnm{Peng}, \binits{W.}},
\bauthor{\bsnm{Park}, \binits{S.Y.}},
\bauthor{\bsnm{Roh}, \binits{C.}},
\bauthor{\bsnm{Mun}, \binits{J.}},
\bauthor{\bsnm{Ju}, \binits{H.}},
\bauthor{\bsnm{Kim}, \binits{J.}},
\bauthor{\bsnm{Ko}, \binits{E.}},
\bauthor{\bsnm{Liang}, \binits{Z.}},
\bauthor{\bsnm{Hahn}, \binits{S.}},
\bauthor{\bsnm{Zhang}, \binits{J.}},
\bauthor{\bsnm{Sanchez}, \binits{A.}},
\bauthor{\bsnm{Walker}, \binits{D.}},
\bauthor{\bsnm{Hindmarsh}, \binits{S.}},
\bauthor{\bsnm{Si}, \binits{L.}},
\bauthor{\bsnm{Jo}, \binits{Y.}},
\bauthor{\bsnm{Jo}, \binits{Y.}},
\bauthor{\bsnm{Kim}, \binits{T.}},
\bauthor{\bsnm{Kim}, \binits{C.}},
\bauthor{\bsnm{Wang}, \binits{L.}}:
\batitle{Flexoelectric polarizing and control of a ferromagnetic metal}.
\bjtitle{Nature Physics}
\bvolume{20},
\bfpage{1}--\blpage{6}
(\byear{2024})
\doiurl{10.1038/s41567-023-02333-8}
\end{barticle}
\endbibitem

%%% 97
\bibitem[\protect\citeauthoryear{Lenthe et~al.}{2019}]{LENTHE2019}
\begin{barticle}
\bauthor{\bsnm{Lenthe}, \binits{W.C.}},
\bauthor{\bsnm{Singh}, \binits{S.}},
\bauthor{\bsnm{Graef}, \binits{M.D.}}:
\batitle{A spherical harmonic transform approach to the indexing of electron back-scattered diffraction patterns}.
\bjtitle{Ultramicroscopy}
\bvolume{207},
\bfpage{112841}
(\byear{2019})
\doiurl{10.1016/j.ultramic.2019.112841}
\end{barticle}
\endbibitem

%%% 98
\bibitem[\protect\citeauthoryear{Boeck et~al.}{2011}]{BOECK2011}
\begin{barticle}
\bauthor{\bsnm{Boeck}, \binits{S.}},
\bauthor{\bsnm{Freysoldt}, \binits{C.}},
\bauthor{\bsnm{Dick}, \binits{A.}},
\bauthor{\bsnm{Ismer}, \binits{L.}},
\bauthor{\bsnm{Neugebauer}, \binits{J.}}:
\batitle{The object-oriented {DFT} program library {S/PHI/nX}}.
\bjtitle{Computer Physics Communications}
\bvolume{182}(\bissue{3}),
\bfpage{543}--\blpage{554}
(\byear{2011})
\doiurl{10.1016/j.cpc.2010.09.016}
\end{barticle}
\endbibitem

%%% 99
\bibitem[\protect\citeauthoryear{Perdew et~al.}{1996}]{perdew1996}
\begin{barticle}
\bauthor{\bsnm{Perdew}, \binits{J.P.}},
\bauthor{\bsnm{Burke}, \binits{K.}},
\bauthor{\bsnm{Ernzerhof}, \binits{M.}}:
\batitle{Generalized gradient approximation made simple}.
\bjtitle{Physical Review Letters}
\bvolume{77},
\bfpage{3865}--\blpage{3868}
(\byear{1996})
\doiurl{10.1103/PhysRevLett.77.3865}
\end{barticle}
\endbibitem

%%% 100
\bibitem[\protect\citeauthoryear{Hachtel et~al.}{2018}]{hachtel2018}
\begin{botherref}
\oauthor{\bsnm{Hachtel}, \binits{J.}},
\oauthor{\bsnm{Idrobo}, \binits{J.}},
\oauthor{\bsnm{Chi}, \binits{M.}}:
Sub-{A}ngstrom electric field measurements on a universal detector in a scanning transmission electron microscope.
Advanced Structural and Chemical Imaging
\textbf{4}
(2018)
\doiurl{10.1186/s40679-018-0059-4}
\end{botherref}
\endbibitem

%%% 101
\bibitem[\protect\citeauthoryear{Müller-Caspary et~al.}{2017}]{MULLERCASPARY2017}
\begin{barticle}
\bauthor{\bsnm{Müller-Caspary}, \binits{K.}},
\bauthor{\bsnm{Krause}, \binits{F.F.}},
\bauthor{\bsnm{Grieb}, \binits{T.}},
\bauthor{\bsnm{Löffler}, \binits{S.}},
\bauthor{\bsnm{Schowalter}, \binits{M.}},
\bauthor{\bsnm{Béché}, \binits{A.}},
\bauthor{\bsnm{Galioit}, \binits{V.}},
\bauthor{\bsnm{Marquardt}, \binits{D.}},
\bauthor{\bsnm{Zweck}, \binits{J.}},
\bauthor{\bsnm{Schattschneider}, \binits{P.}},
\bauthor{\bsnm{Verbeeck}, \binits{J.}},
\bauthor{\bsnm{Rosenauer}, \binits{A.}}:
\batitle{Measurement of atomic electric fields and charge densities from average momentum transfers using scanning transmission electron microscopy}.
\bjtitle{Ultramicroscopy}
\bvolume{178},
\bfpage{62}--\blpage{80}
(\byear{2017})
\doiurl{10.1016/j.ultramic.2016.05.004} .
\bcomment{FEMMS 2015}
\end{barticle}
\endbibitem

%%% 102
\bibitem[\protect\citeauthoryear{Chapman et~al.}{1978}]{CHAPMAN1978}
\begin{barticle}
\bauthor{\bsnm{Chapman}, \binits{J.N.}},
\bauthor{\bsnm{Batson}, \binits{P.E.}},
\bauthor{\bsnm{Waddell}, \binits{E.M.}},
\bauthor{\bsnm{Ferrier}, \binits{R.P.}}:
\batitle{The direct determination of magnetic domain wall profiles by differential phase contrast electron microscopy}.
\bjtitle{Ultramicroscopy}
\bvolume{3},
\bfpage{203}--\blpage{214}
(\byear{1978})
\doiurl{10.1016/S0304-3991(78)80027-8}
\end{barticle}
\endbibitem

%%% 103
\bibitem[\protect\citeauthoryear{Gao et~al.}{2019}]{gao2019}
\begin{barticle}
\bauthor{\bsnm{Gao}, \binits{W.}},
\bauthor{\bsnm{Addiego}, \binits{C.}},
\bauthor{\bsnm{Wang}, \binits{H.}},
\bauthor{\bsnm{Yan}, \binits{X.}},
\bauthor{\bsnm{Hou}, \binits{Y.}},
\bauthor{\bsnm{Ji}, \binits{D.}},
\bauthor{\bsnm{Heikes}, \binits{C.}},
\bauthor{\bsnm{Zhang}, \binits{Y.}},
\bauthor{\bsnm{Li}, \binits{L.}},
\bauthor{\bsnm{Huyan}, \binits{H.}},
\bauthor{\bsnm{Blum}, \binits{T.}},
\bauthor{\bsnm{Aoki}, \binits{T.}},
\bauthor{\bsnm{Nie}, \binits{Y.}},
\bauthor{\bsnm{Schlom}, \binits{D.}},
\bauthor{\bsnm{Wu}, \binits{R.}},
\bauthor{\bsnm{Pan}, \binits{X.}}:
\batitle{Real-space charge-density imaging with sub-ångstrom resolution by four-dimensional electron microscopy}.
\bjtitle{Nature}
\bvolume{575},
\bfpage{1}--\blpage{1}
(\byear{2019})
\doiurl{10.1038/s41586-019-1649-6}
\end{barticle}
\endbibitem

%%% 104
\bibitem[\protect\citeauthoryear{Müller-Caspary et~al.}{2014}]{mullercaspary2014}
\begin{barticle}
\bauthor{\bsnm{Müller-Caspary}, \binits{K.}},
\bauthor{\bsnm{Krause}, \binits{F.}},
\bauthor{\bsnm{Béché}, \binits{A.}},
\bauthor{\bsnm{Schowalter}, \binits{M.}},
\bauthor{\bsnm{Galioit}, \binits{V.}},
\bauthor{\bsnm{Löffler}, \binits{S.}},
\bauthor{\bsnm{Verbeeck}, \binits{J.}},
\bauthor{\bsnm{Zweck}, \binits{J.}},
\bauthor{\bsnm{Schattschneider}, \binits{P.}},
\bauthor{\bsnm{Rosenauer}, \binits{A.}}:
\batitle{Atomic electric fields revealed by a quantum mechanical approach to electron picodiffraction}.
\bjtitle{Nature Communications}
\bvolume{5},
\bfpage{5653}
(\byear{2014})
\doiurl{10.1038/ncomms6653}
\end{barticle}
\endbibitem

%%% 105
\bibitem[\protect\citeauthoryear{Shibata et~al.}{2012}]{shibata2012}
\begin{barticle}
\bauthor{\bsnm{Shibata}, \binits{N.}},
\bauthor{\bsnm{Findlay}, \binits{S.}},
\bauthor{\bsnm{Kohno}, \binits{Y.}},
\bauthor{\bsnm{Sawada}, \binits{H.}},
\bauthor{\bsnm{Kondo}, \binits{Y.}},
\bauthor{\bsnm{Ikuhara}, \binits{Y.}}:
\batitle{Differential phase-contrast microscopy at atomic resolution}.
\bjtitle{Nature Physics}
\bvolume{8},
\bfpage{611}--\blpage{615}
(\byear{2012})
\doiurl{10.1038/nphys2337}
\end{barticle}
\endbibitem

%%% 106
\bibitem[\protect\citeauthoryear{P.}{1927}]{ehrenfest1927}
\begin{barticle}
\bauthor{\bsnm{P.}, \binits{E.}}:
\batitle{Bemerkung ueber die angenaeherte,{G}ueltigkeit der klassischen {M}echanik innerhalb der {Q}uantenmechanik}.
\bjtitle{Zeitschrift für Physik}
(\byear{1927})
\doiurl{10.1007/BF01329203}
\end{barticle}
\endbibitem

%%% 107
\bibitem[\protect\citeauthoryear{Pennycook~SJ}{2011}]{pennycook2011}
\begin{bbook}
\bauthor{\bsnm{Pennycook~SJ}, \binits{E.} \bsuffix{Nellist~PD}}:
\bbtitle{Scanning Transmission Electron Microscopy}.
\bpublisher{Springer},
\blocation{New York}
(\byear{2011}).
\doiurl{10.1007/978-1-4419-7200-2}
\end{bbook}
\endbibitem

\end{thebibliography}

\end{document}